\documentclass[a4paper,fleqn,usenatbib]{mnras}

\usepackage{newtxtext,newtxmath,longtable}
\usepackage{epstopdf}
\usepackage{longtable}
\usepackage{lscape} 
\usepackage{graphicx,longtable,lscape,natbib,amssymb,amssymb,amsmath}
\usepackage{ragged2e}

\usepackage[T1]{fontenc}
\usepackage{ae,aecompl}

\usepackage{graphicx}	
\usepackage{amsmath}	
\usepackage{amssymb}	

\title[The Te-REX sample]{Te-REX: a sample of extragalactic TeV-emitting candidates}

\author[B. Balmaverde et al.]{
Barbara Balmaverde,$^{1,2}$\thanks{E-mail: barbara.balmaverde@inaf.it}
A. Caccianiga$^{2}$,
R. Della Ceca$^{2}$,
A. Wolter$^{2}$,
A. Belfiore$^{3}$,
 \newauthor
 L. Ballo$^{4}$,
M. Berton$^{5,6}$,
I. Gioia$^{7}$,
T. Maccacaro$^{2}$,
and B. Sbarufatti$^{8}$
\\
\\
$^{1}$INAF - Osservatorio Astrofisico di Torino, Via Osservatorio 20, 10025 Pino Torinese, Italy\\
$^{2}$INAF - Osservatorio Astronomico di Brera, Via Brera 28, 20121 Milano, Italy\\
$^{3}$INAF/IASF Milano, Via Alfonso Corti 12, I-20133 Milano, Italy\\
$^{4}$XMM-Newton Science Operations Centre, ESAC/ESA, PO Box 78, E-28691 Villanueva de la can\~{a}da, Madrid, Spain\\
$^{5}$ Finnish Centre for Astronomy with ESO (FINCA), University of Turku, Quantum, Vesilinnantie 5, FIN-20014 Turku, Finland\\
$^{6}$Aalto University Mets\"ahovi Radio Observatory, Mets\"ahovintie 114, FIN-02540 Kylm\"al\"a, Finland\\
$^{7}$INAF-Istituto di Radioastronomia, via Gobetti 101, I-40129 Bologna, Italy\\
$^{8}$Department of Astronomy and Astrophysics, Pennsylvania State University. 525 Davey Laboratory, University Park, PA 16802, USA\\
}

\date{Accepted XXX. Received YYY; in original form ZZZ}

\pubyear{2019}

\begin{document}
\label{firstpage}
\pagerange{\pageref{firstpage}--\pageref{lastpage}}
\maketitle

\begin{abstract}
\\

The REX (Radio-Emitting X-ray sources) is a  catalogue produced by cross-matching X-ray data from the ROSAT-PSPC archive of pointed observations and radio data from the NRAO VLA Sky Survey, aimed at the selection of  blazars. 
From the REX catalogue,  we select a well defined and statistically complete sample  of  high-energy peaked BL Lac (HBL). HBL are  expected to be the most numerous class of
extragalactic TeV emitting sources. 
Specifically,  we have considered only the REX sources in the currently planned CTA extragalactic survey area satisfying specific criteria and with an optical spectroscopic confirmation.
 We obtain 46 HBL candidates  that we called Te-REX ({\it TeV-emitting REX}).
We estimate the very high-energy gamma-ray emission, in the TeV domain, using an empirical approach i.e. 
using specific statistical relations between gamma-rays (at GeV energies) and radio/X-rays properties observed in bright HBL  from the literature. 
We compare the spectral energy distributions (SEDs) with the sensitivities of current and upcoming 
Cherenkov telescopes  and we predict that  14 Te-REX could be detectable with 50 hours of observations of CTA and 7 of them also with current Cherenkov facilities in 50 hours.  By extrapolating these numbers on the total extragalactic sky, we predict that about 800 HBL could be visible in pointed CTA observations and $\sim$400 with current Cherenkov telescopes in 50 hours. Interestingly, our predictions show that a non-negligible fraction ($\sim$30\%) of the HBL that will be detectable by CTA is composed of relatively weak objects whose optical nuclear emission is swamped by the host-galaxy light and not (yet) detected by Fermi-LAT.
\end{abstract}




\section{Introduction}

\label{intro}

The extragalactic gamma-ray sky is dominated by blazars, 
a sub-class of radio loud Active Galactic Nuclei (AGN) having their relativistic jet pointing towards our line of sight \citep{blandford78}. Their spectral energy distribution (SED)
 is almost completely produced by non-thermal processes: synchrotron emission from relativistic electrons and Inverse Compton emission, e.g., \citep{padovani95,fossati98,ghisellini98}. 
  In the lepto-hadronic scenarios,  both synchrotron and inverse Compton processes contribute significantly to the high energy bump. These models well reproduce the broadband SEDs of blazars (e.g. \citealt{cerruti11}).
  Blazars in which the synchrotron bump peaks at very high frequency (up to X-rays, Log$\nu_{p}^{S}$(Hz)$>$15.5; \citealt{fan16}) are called High-Energy peaked BL Lac (HBL) and are the
  most numerous class of extragalactic sources detected by the current generation of Cherenkov telescopes \citep{massaro11}. HBLs are 
  the most promising candidates, among the extragalactic sources, to be detected by the upcoming Cherenkov Telescope Array (CTA) that will improve the sensitivity of current 
  Imaging Air Cherenkov telescopes (IACTs) by an order of magnitude.
To put all the investigations on a statistical basis it is fundamental to have an unbiased and complete sample of HBL that thus can be used to select good candidates for observation with the current and upcoming Cherenkov telescopes, like the Cherenkov Telescope Array (CTA, \citealt{science}) and ASTRI (Astrofisica con Specchi a Tecnologia replicante Italiana, \citealt{astri}).
 
Recently, a large sample of HBL candidates has been published (the 2WHSP catalogue, \citealt{chang17}) containing $\sim$1700 sources obtained by combining different radio, infrared, X-ray and gamma-ray (Fermi-LAT) catalogues. This sample is expected to contain many of the HBL that will be  detected in the near future by the new generation of Cherenkov telescopes. However, due to the type of selection which was essentially based on photometric criteria, this sample is not expected to be complete and representative of all the HBL population, in a given area of sky and down to a well defined flux limit. For instance, if the optical/IR nuclear emission of a low-luminosity HBL is heavily dominated by the host galaxy light, the usual photometric criteria (e.g., \citealt{massaro12}) are not expected to work. Our goal is to create a complete and independent sample,  more representative of the entire HBL population.

To this  aim, we started from a
well defined sample of  Radio-Emitting X-ray sources (REX, \citealt{caccianiga99}, CA99 in the following) that was selected by 
combining X-ray and radio
observations  from ROSAT-PSPC (Position Sensitve Proportional Counter, \citealt{voges96,white94}) and the NRAO VLA Sky Survey (NVSS, \citealt{condon98}).  
These two datasets, 
 thanks to the low flux limits in the radio (5 mJy at 1.4 GHz) and in the X-ray band ($\sim$5$\times$10$^{-14}$ ergs s$^{-1}$ cm$^{-2}$, in the 0.5-2.0 keV band) 
 and the large area of sky covered, provide an optimal compromise between
 coverage and depth and they are  suitable for selecting an unbiased sample of  HBLs.
Up to now the REX
catalogue still represents an excellent starting point for an unbiased search of HBLs objects on a relatively wide sky area. 

In this paper we present a sub-sample of HBL selected from the REX survey. We call it the Te-REX sample (TeV-emitting REX candidates), i.e. REX  that are possible TeV-emitters. 
Since our selection does not rely on existing gamma-ray (e.g., Fermi-LAT) catalogues and since it does not apply filters on the IR/optical colours of the sources, we expect that the Te-REX sample can provide a representative view of the HBL population that dominates the very high energies (VHE) sky. 

The paper is organised as follows: in Section 2 we review the REX project, in Section 3 we describe the  criteria used to select a sub-sample 
of TeV-emitting candidates (Te-REX), in Section 4 we discuss the  properties of this sample and we compare them
with other high energy sources catalogues. In Section 5 we extrapolate the SED at  VHE and in Section 6 we compare these SED with the sensitivity of current and future Cherenkov telescopes. In Section 7 we discuss our results  and finally we present our summary and conclusions  in Section 8.

\section{The REX project}
\label{results}

BL Lacs are difficult objects
 to find, because  they lack both
 a UV excess and any significant spectral features (EW$<$5\AA, \citealt{stocke91,perlman98}); moreover, they represent only a  few percent of the total AGN population. 
 Nevertheless, they are both
  radio and X-ray-loud (e.g. \citealt{stocke90}) and therefore they are traditionally selected either in the X-rays or in the radio band.
  In the recent years, the gamma-ray windows has been proven to be very effective in selecting BL Lac objects, thanks also to the Fermi-LAT telescope. 
However, since the positional uncertainty of the FERMI sources does not allow to unambiguously pin-point the optical counterpart, the search for the counterpart is usually based on a radio/X-ray detection (or a combination of the two). 
  
The REX project has the main scientific goal of selecting 
 a large and statistically representative sample of BL Lac object. The REX is a sample of $\sim$ 1600 radio-emitting X-ray sources selected through 
 a positional cross-correlation of data from a VLA survey (NVSS) and the archival ROSAT PSPC pointed observations (each one covering about 2.8 deg$^2$) covering a large area of the sky (more than $\approx$2000 deg$^2$, CA99).
 The REX catalogue was built using only the ROSAT PSPC pointing satisfying the criteria listed in CA99, having excluded
in each field the target of the observation, to avoid any possible bias towards particular classes of objects (see CA99 for more details).  The accuracy of the radio/X-ray  association is  limited by the X-ray spatial uncertainties, typically of the order of tens of arc-seconds (from 14$"$ in the center of the ROSAT PSPC to about 60$"$ in the outermost portion of the ROSAT PSPC detector), while the error circles associated with the NVSS positions are an order of magnitude smaller (the 90\% confidence error on the radio source position  is typically smaller than 5$"$). The cross-correlation resulted  in a catalogue with a 90\% completeness level and with 10\% of expected spurious sources.

  We stress that to define  the REX sample
  we do not impose any particular preselection criteria 
 except for the presence of the source in a radio and in an X-ray catalogue. 
The positional accuracy of the VLA data  
 guarantees that there is in general only one possible optical counterpart, at least for bright magnitudes ($<$21). The optical counterparts for most of the REX sources have been recently refined  using the 
Panoramic Survey Telescope and Rapid Response System (Pan-STARRS, \citealt{chambers16}) survey\footnote{The Pan-STARRS survey
used a 1.8 meter telescope in Hawaii to image the sky in five broadband  filters (g, r, i, z, y).}.

 
 
Several observing runs at different telescopes have been performed to identify the REX sources (see CA99,\citealt{wolter98,caccianiga00} and \citealt{caccianiga02}).
These optical spectroscopic observations have been carried out during the period 1998/2002 and 2018 at
 the UH 88$"$ telescope  at Mauna Kea (USA), at UNAM 2.1m at  San Pedro Martir (Mexico), at ESO 3.6m, 2.2m and 1.5m telescopes in 
 La Silla, (Chile) and at the Telescopio Nazionale Galileo (TNG) in La Palma (Spain), leading to the classification of about half of the REX sources.


 

\begin{table*}
\caption{Journal of the observations}
\label{sample}
\begin{tabular}{l l l l l l}
\hline\hline
Telescope$^1$/Instrument$^2$ & Grism name  & Dispersion & Observing Period & REF \\
   & &  [\AA/pixel]     & & \\
\hline
UH 88''+ WFGS & 400  & 4.1 & 1998 Feb 26 -- Mar 1 & MK2/98\\
UH 88''+ WFGS & 400  & 4.1 & 1999 Feb 22 -- 24 & MK2/99\\
UH 88''+ WFGS & 400  & 4.1 & 1999 Mar 24 -- 25 & MK3/99\\
INT+ IDS           & R300V &  3.3 & 1999 May 7 -- 12 & INT5/99\\
CA 2.2m+CAFOS  & B200  & 4.5 & 1999 Jun 29 -- Jul 6 & CA7/99\\
UH 88''+ WFGS & 400  & 4.1 & 2000 Feb 10 -- 13 & MK2/00\\
UH 88''+ WFGS & 400  & 4.1 & 2001 Mar 16 -- 19 & MK3/01\\
UH 88''+ WFGS & 400  & 4.1 & 2002 Apr  11 -- 13 & MK4/02\\
ESO 3.6m + EFOSC2  & Gr.6  & 2.1  & 2002 May 4 -- 7 & ESO5/02a\\  
ESO 1.5m + B\&C & Gr.15 &  3.8  & 2002 May 2 -- 3 & ESO5/02b \\
TNG+DOLORES  & LR-B  & 2.5 & 2002 Sep 9 -- 12 & TNG9/02\\
TNG+DOLORES  & LR-B  & 2.5 & 2018 May 16 -- 19 & TNG5/18\\
\hline
\end{tabular}
\\
\justify{
$^1$ {\bf Telescopes:}  UH 88'' = 2.2 meter telescope of the University of Hawaii, located at Mauna Kea (US); 
INT = Isaac Newton Telescope, located at La Palma (Spain); 
CA 2.2m = 2.2 meter telescope at the Calar Alto Observatory located in Almeria (Spain); 
ESO 3.6m = 3.6 meter telescope of the European Southern Observatory located at La Silla (Chile); 
ESO 1.5m = 1.52 meter telescope of the European Southern Observatory located at La Silla (Chile); 
TNG = Telescopio Nazionale Galileo located at La Palma (Spain);
$^2$ {\bf Instruments}: WFGS = Wide Field Grism Spectrograph;
IDS = Intermediate Dispersion Spectrograph;
CAFOS = Calar Alto Faint Object Spectrograph;
EFOSC2 = ESO Faint Object Spectrograph and Camera (v.2);
B\&C = Boller and Chivens Spectrograph;
DOLORES = Device Optimized for the LOw RESolution;
{\bf REF:} Reference code for the spectroscopical run identified with the telescope name, month and year of observation (this code will be used in Tab. B1).
{\bf Note:} in all cases, a long-slit was used with a width ranging from 1 to 1.5 arcsec depending 
on the seeing conditions.}
\label{logs}
\end{table*}

  \section{From the REX to the Te-REX sample}
\label{sel}

In order to select a well defined and unbiased sample of HBL we consider only the objects in the REX survey with the highest X-ray-to-radio flux ratio, since there is a relatively strict correlation between this ratio and the position of the synchrotron peak, as discussed in Section~5.1.
In particular, we apply a selection on the basis of the two-point spectral index\footnote{We define the two-point spectral index  between 5 GHz and 1 keV as
 $\alpha_{\rm RX}=-Log(S_{5 GHz}/S_{1keV})/ Log(\nu_{\rm 5 GHz}/\nu_{\rm 1keV}$) where $S_{\rm 5 GHz}$ and $S_{\rm 1keV}$ are the monochromatic flux densities defined at 5 GHz and 1 keV respectively and $\nu_{\rm 5 GHz}$ and $\nu_{\rm 1keV}$ are the corresponding frequencies. 
 For the k-corrections and the conversion from the observed X-ray fluxes to the monochromatic fluxes we have assumed $\alpha_r= 0$ and  $\alpha_X= 1$.} $\alpha_{\rm RX}$. Since  high-energy peaked sources are characterised by small (flat) values of the radio-to-X-ray spectral index \citep{padovani95}, 
we selected sources with $\alpha_{\rm RX} \le$0.74.  We show in Section \ref{stat} that this criterium selects sources with a logarithm of the frequency of the synchrotron peak larger than 15.5, usually classified as HBL \citep{fan16}.

In principle, a high X-ray-to-radio flux ratio could also lead to the selection of clusters/groups of galaxies, if they contain a radio galaxy. In these cases
the X-ray emission is produced by  the hot diffuse plasma and it is unrelated  to the radio emission. Radio galaxies in clusters are thus expected to contaminate the selection of HBL objects. To reduce this contamination, we discard the extended sources in the 
ROSAT PSPC image i.e. sources with an X-ray size larger by 20\% than the local PSF.



In addition, we impose a limit on the optical magnitude of the counterpart (mag$_i<$21) to  maximise the level of spectroscopic identification of the sample. We note that this limit does not exclude optically weak BL Lacs, at least at redshift below 0.5-0.6, thanks to the presence of the host galaxy whose optical magnitude is expected to be brighter than i=21 for z$<$0.6 (e.g. \citealt{Sbarufatti05}). For higher redshifts, instead, the magnitude limit may affect the completeness of the sample. However, we do not expect significant TeV emission from BL Lac at z$>$0.6 due the effect of the extragalactic background light (EBL) absorption \citep{gould67,vassiliev00} and, therefore, we expect that the resulting sample will be reasonably complete for what concerns potentially detectable TeV emitters.

Finally, we have chosen a specific sky area of 560 deg$^2$, included in the one currently planned for the CTA extragalactic survey \citep{science}.

 
 Summarising, we applied the following selection criteria to the sources:
 
\begin{itemize}
 \item Galactic coordinates in the sky region defined by the following constraints b$^{\rm{II}}\ge$10$^\circ$, l$^{\rm{II}}\ge$270$^\circ$ or l$^{\rm{II}}\le$90$^\circ$ 
 \item  X-ray size $<$ 1.2 times the PSPC PSF size at the source position 
 \item mag$_i\le$21 (from the Pan-STARRS catalogue)
 \item $\alpha_{\rm RX}\le$ 0.74

\end{itemize}

These criteria define a list of 87 REX sources that are listed in Table  \ref{sampleall}.
 Even if this sample has been selected using a radio and an X-ray flux limit, it can be considered as an (almost) purely radio selected sample for what concerns the HBL candidates. Indeed, due to their high X-ray-to-radio flux ratio virtually all the sources in the sample that make the radio flux limit are also expected to be detected above the X-ray flux limit. We stress again that the selection criteria are independent from any source properties in the GeV-TeV band.

\begin{figure*}  
\centering{ 
\includegraphics[width=6.5cm,angle=-90]{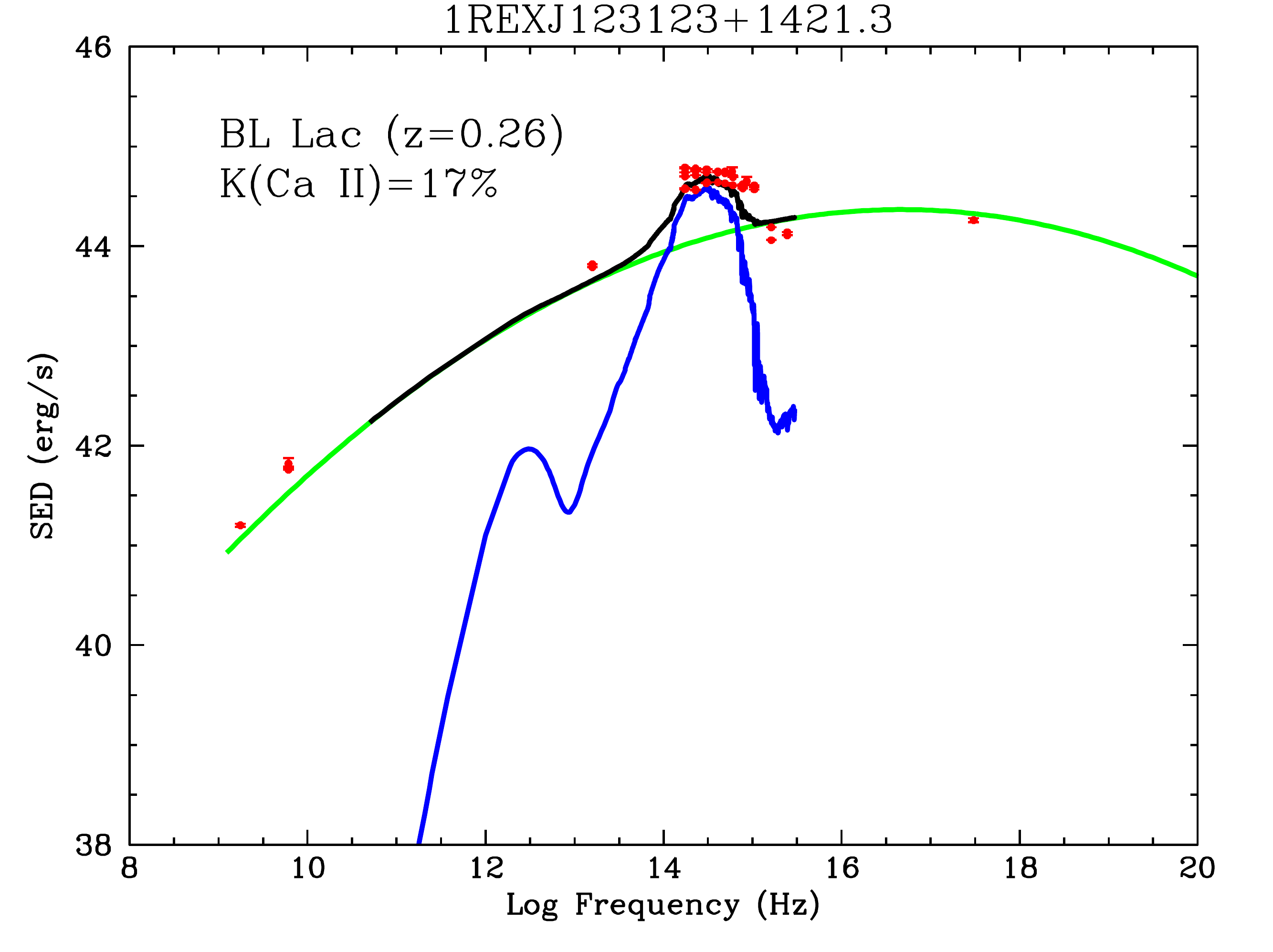} 
\includegraphics[width=6.5cm,angle=-90]{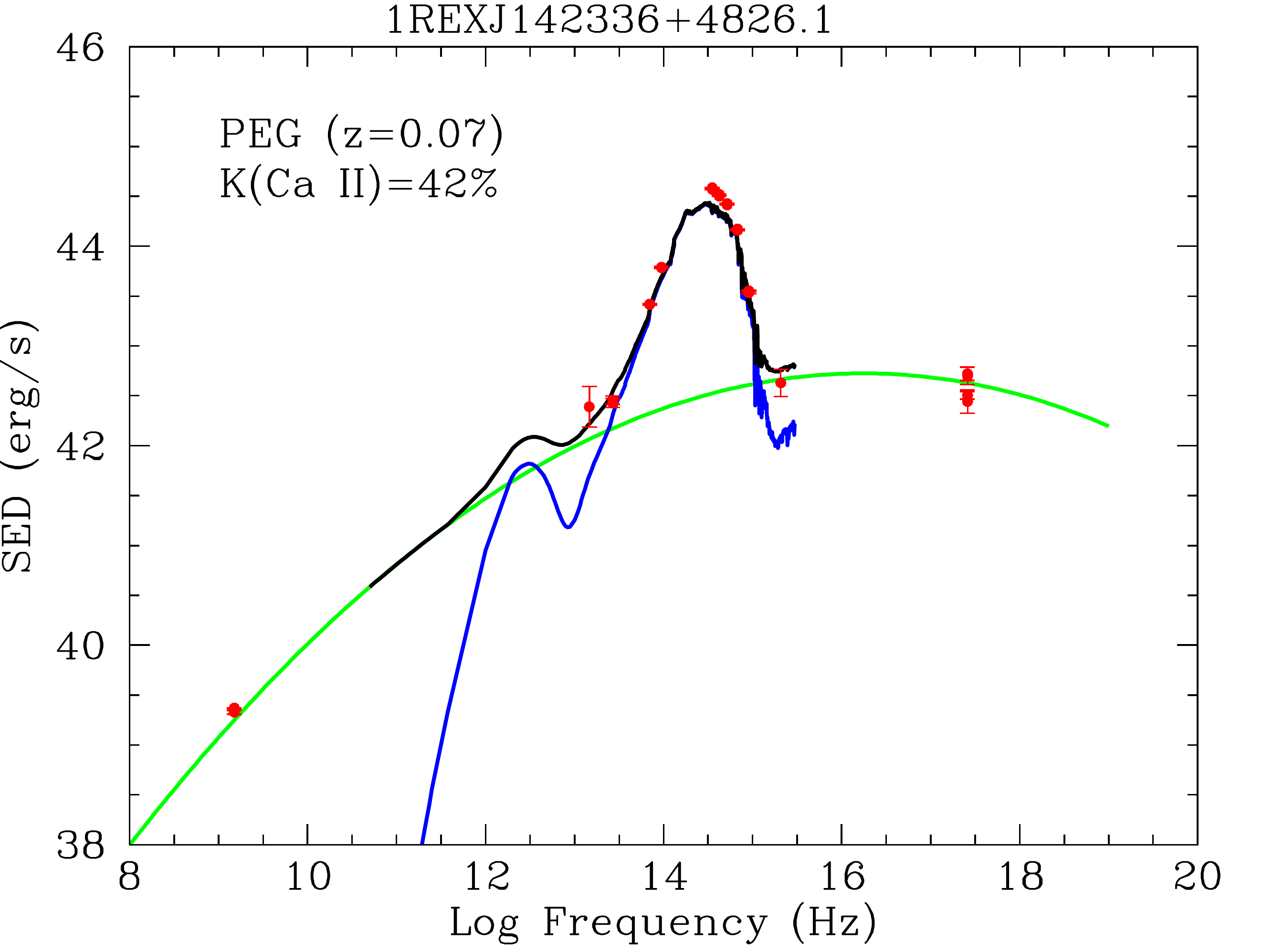} 
\caption{Example of the SEDs of two Te-REX sources, one spectroscopically classified as BL Lac (left:1REXJ123123+1421.3) and one as PEG (right:1REXJ142336+4826.1) showing different ratios of thermal to non-thermal components: in the source on the left side the non-thermal continuum (green line) is high compared to the host galaxy (in blue, template from SSDC SED builder) and, therefore, the BL Lac nucleus is clearly detected in the optical spectrum as a significant reduction of the CaII break (K(CaII)$\sim$17\%); the source on the right side, instead, is very similar in terms of non-thermal shape (the position of the synchrotron peak frequency is very similar) but has a much lower  intensity. Therefore the host galaxy (which is very similar in intensity to the previous case) is now dominating the optical spectrum giving no chances to detect the BL Lac nucleus spectroscopically. We classify this object as "PEG" indicating that this source might contain  a low-luminosity BL Lac. Clearly sources like these are  typically selected at low redshifts, given their low luminosity, and, therefore, they are good candidates for the detection at VHE.  Data points (in red) are derived from the SSDC SED builder tool.}
\label{ssdc}}
\end{figure*}  

 \subsection{Spectroscopic  classification of the Te-REX sample}
\label{sample}

For many targets we had spectroscopic data from our own observing runs
 or from
the Sloan Digital Sky Survey (SDSS, \citealt{stoughton02}) archive.  
In an observing run carried out at the TNG telescope in may 2018 we almost completed the spectroscopic identification of this Te-REX sample (up to 98\%) leaving only
  2 sources unclassified. 
 We analyzed all the optical spectra, including those collected in the past, to provide a uniform classification.
 For the data reduction we have used the IRAF longslit package. In Table \ref{logs} we provide the journal of observations.

We classify as E.L.AGN (Emission Line AGN) or E.L.G (Emission line galaxies) sources with strong emission lines (EW$\ge$5\AA) and we do not consider those objects further in this paper. For the spectra with no or weak emission lines (EW$<$5\AA) we evaluate the relative depression of the continuum across $\lambda$=4000\AA, due to the absorption lines of ionised metals in the atmosphere of stars, that becomes larger for older and metal-richer stellar population. 
We have computed its amplitude following \citet{bruzual83}, estimating D$_n$(4000), the ratio of the average flux densities (expressed in units of frequency) between 4050\AA\, and 4250\AA\, and between 3750\AA\, and 3950\AA\,
 in the rest frame of the source. We define K(CaII), known as "Calcium II break", as 
 \begin{equation}
 K({\rm Ca II}) =(1-1/D_n(4000))\times100,
 \end{equation}
 the  decrease of the flux (in percentage) between the blue and red 
 portion of the spectrum around 4000\AA. In a normal early type galaxy, the value of K(CaII) is close to 50\%, with a typical range between 40\% and 60\%. If an additional non-thermal emission  is present, implying the presence of an active nucleus, the value of K decreases.  The stronger the non-thermal emission, the lower is the value of K(CaII). 
  This method  to classify a source as BL Lac object has been proposed  by \citet{stocke91} and has been widely applied in literature. Following CA99
 we classify an object as BL Lac if there are no strong emissione lines (EW$<$5 \AA) and if the Calcium II break is below 40\%. 
When the non-thermal emission is dominant, i.e. we observe a completely featureless spectrum, the value of K(CaII) depends only on the slope of the nuclear 
emission  and can reach negative values, for very blue nuclear spectra. If we cannot measure the redshift from the spectrum, in the following analysis we  assume a redshift value of z=0.3 (six sources), the median value obtained from the sources with measured redshifts. 

When K(CaII) is above 40\%, instead, the star-light emission is dominant and we cannot establish the presence of an active nucleus from the optical spectrum. We classify these sources as Passive Elliptical Galaxies (PEGs, \citealt{marcha01}).
As explained in the following section, we believe that  the low-luminosity tail of the BL Lac population is made mostly by PEGs.  Keeping the PEGs as possible BL Lac is a conservative choice also because nuclear variability can move one source from one class to another (PEG or BL Lac).
In Tab. \ref{sampleall} we list the 87 REX sources with the optical spectroscopic classification as emission line object  (emission line AGN, E.L.AGNs, or emission line galaxy, E.L.G.), passive elliptical galaxies (PEGs) or BL Lacs Objects (BL). In Fig.\ref{spectraall} we present the spectra for all the sources (but two\footnote{ The two sources with non reported spectra are
 1REXJ133529-2950.6
and 1REXJ184120+5906.1. The identification is taken from the literature and no optical spectrum in electronic form is available to us.}) classified as PEGs (21 objects) or BL Lacs (25 objects). 

\subsection{Are Te-REX PEGs really low-luminosity BL Lacs?}
 
The traditional criteria for selecting BL Lacs, based on e.g. photometric colours
or properties of the optical spectra,
might systematically miss low luminosity BL Lacs. In fact most of them are expected to have the optical/IR nuclear emission heavily diluted by the host galaxy light and, therefore, their optical spectra are  likely indistinguishable from those of normal, non active-galaxies. 
We have found in the Te-REX sample a significant number of sources with no obvious sign of nuclear emission in the optical spectrum. As explained in the previous section, we classify these objects as PEGs. These sources are characterised by an 
optical spectrum without strong emission lines and a compact X-ray emission coincident with the radio emission. They have a similar  $\alpha_{\rm RX}$ (i.e. possible similar synchrotron peak) with respect to classical HBL but different radio power ($\sim10^{23}$ vs 10$^{25}$ W/Hz).  Clearly, since blazars are variable objects, the classification as BL Lacs or PEG could change with time.

In Fig.~\ref{ssdc}, we compare the observed SEDs, from radio to X-rays, of two Te-REX classified as BL Lac and PEG, respectively. Both sources have a similar value of  $\alpha_{\rm RX}$ (i.e. a similar synchrotron peak) but a significantly different nuclear luminosity. In the first case the nuclear luminosity is so high that the host galaxy is barely visible in the spectrum, and we classify it as BL Lac (K(${\rm Ca II})<40\%$). In the second case, instead, the nuclear luminosity is two orders of magnitude lower and the host galaxy dominates the optical emission (hence K(${\rm Ca II})>40$\%) leading to the PEG classification.

 As suggested by \citet{marcha13} (see their Fig. 4.), PEGs could represent the "missing" population at low radio luminosities ($\sim$10$^{23}$-10$^{24}$ W Hz$^{-1}$) of beamed up cores of low luminosity radio galaxies (FRI). 
 The  radio luminosity function obtained from \citet{urry91}, considering the beaming effect on the luminosity function of FRI, predicts BL Lacs at the lowest luminosities that are actually not found in samples using the classical definition. Instead, the space density of PEGs is fully consistent with the predictions of the beaming model at the faint end of the radio luminosity function.
 This is a convincing indication that the usual (classical) BL Lac selection simply does not work in this range of lower radio luminosities, and that PEGs could host a blazar nucleus.
 
X-ray data (from Chandra or XMM-Newton) or optical spectra at high resolution could confirm the presence of a blazar in the nucleus of the PEGs. 
However, the clearest indication of blazar nature would came from a detection at TeV energies. Since these sources are preferentially found at low redshift, and therefore are less affected by EBL, CTA will be fundamental in this respect and should be able to detect at least some of them (see Section 6). 

 \begin{figure}  
\centering{ 
\includegraphics[width=9cm]{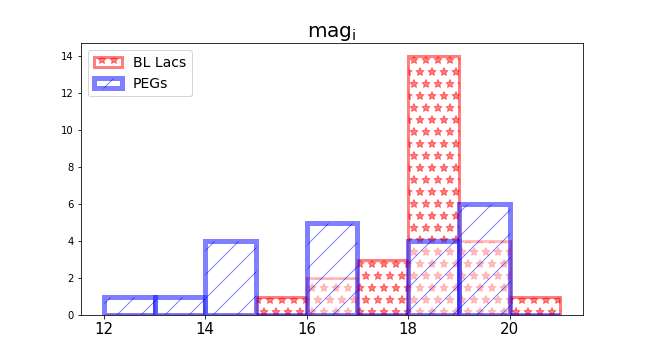} 
\includegraphics[width=9cm]{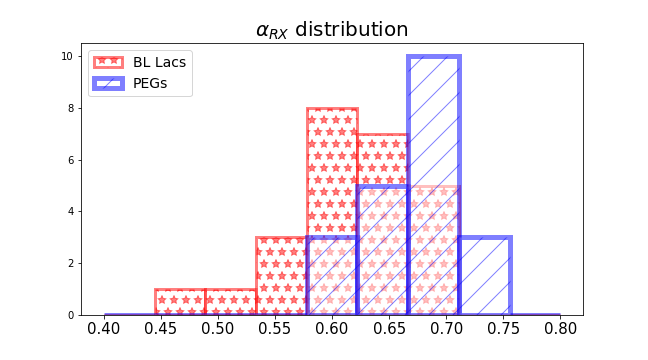} 
\includegraphics[width=9cm]{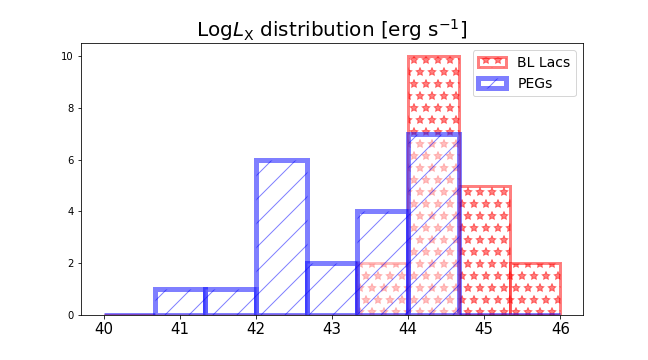} 
\includegraphics[width=9cm]{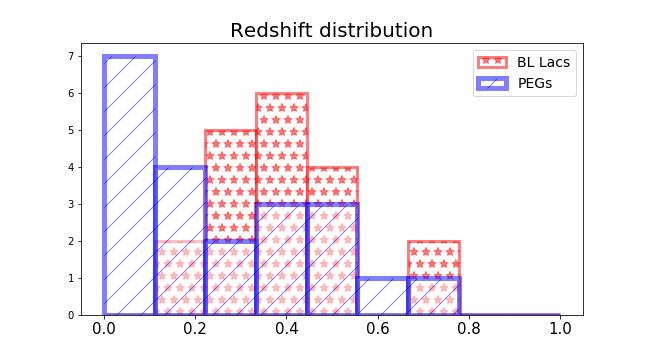} 
\caption{From top to bottom: histograms of the  distribution of the i-band magnitudes, of the 
$\alpha_{\rm RX}$ indexes, of the X-ray luminosities and of the redshifts for the Te-REX sample. Objects classified as PEG and
BL Lacs (HBL candidates) are represented in blue and red colours respectively. }
\label{hist}}
\end{figure} 

\begin{figure}  
\centering{ 
\includegraphics[width=8.5cm]{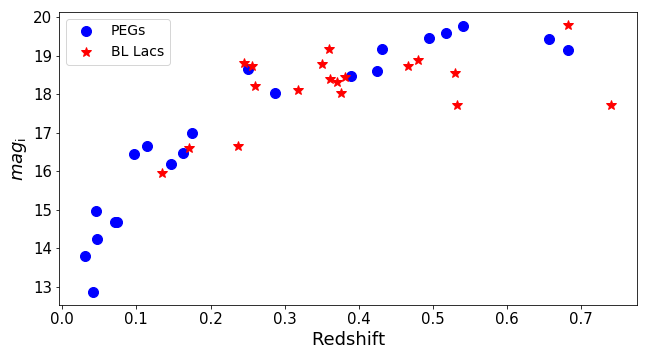} 
\caption{i-band magnitude versus redshift for the sources in the Te-REX sample. Objects classified as PEGs are represented with blue circles and BL Lacs with red stars.}
\label{red}}
\end{figure} 

\section{Properties of the Te-REX sample}

 In the following we refer to the Te-REX  as the sample formed by the 25 BL Lacs and the 21 PEGs (all HBL candidates). 
 The properties of the objects that we have classified as E.L.AGN 
 (likely radio quasar) or E.L.G. (mostly star forming galaxies) will not be investigated in this paper.
 In Fig.~\ref{hist} we explore the properties of the Te-REX considering the 
distribution of the optical i-band magnitudes,  $\alpha_{\rm RX}$ indexes, X-ray luminosities and redshifts.
The host galaxies of BL Lacs  are massive and bright passive ellipticals  \citep{kotilainen98}. The distribution of the apparent i-band magnitudes for the entire sample ranges from mag$_i$=13 to mag$_i$=21.
 However, we note that, with respect to PEGs, BL Lacs  have preferentially fainter magnitudes: PEGs are low luminosity blazar candidates that can be revealed only at low redshifts and hence they typically have a brighter apparent optical magnitude (top panel, Fig. \ref{hist}). 
  Instead BL Lacs are preferentially found in higher redshift galaxies.
BL Lacs are also the objects with the flattest values of $\alpha_{\rm RX}$ and typically have higher X-ray luminosity 
with respect to PEG galaxies (second and third panel from the top, Fig. \ref{hist}). The most likely explanation for this difference is that flat values of $\alpha_{\rm RX}$ correspond also to higher optical nuclear emission and, therefore, to higher nuclear-to-host galaxy luminosity ratio compared to steep $\alpha_{\rm RX}$ values. As a consequence, the objects with a flat $\alpha_{\rm RX}$ have higher chances to be recognised as BL Lac compared to objects with steeper $\alpha_{\rm RX}$ and similar radio emission.

As described in Section~\ref{sel},  our selection criterium (mag$_i<21$) results in a narrow distribution in the redshift parameter space (bottom panel, Fig. \ref{hist}). 
 The redshift histogram for PEG and BL Lacs is centred around z$\sim$0.3. However, the lack of BL Lacs  at redshift z$>$0.8 could be a selection effect related to the fact that in high-z (hence high-luminosity) BL Lacs the non-thermal nucleus likely dominates the optical spectrum making the redshift determination difficult if not impossible. 
 
 In Table~\ref{smalltab} we report the median parameters values for the two different sub-samples.

\begin{table}
\caption{Median parameters values for  passive elliptical galaxies (PEGs) and BL Lacs. Apparent and absolute magnitudes are in the i-band.}
\centering{
\begin{tabular}{lcc}
 & PEG & BL Lacs \\
\hline
mag$\rm _i$ & 17.0 & 18.4 \\
Mag$\rm _i$ & -22.7 & -22.9 \\
Redshift & 0.27 & 0.36 \\
$\alpha_{\rm RX}$ & 0.67 & 0.62 \\
L$_X$ [erg$\,$s$^{-1}$] & 9.0$\times10^{43}$ & 5.3$\times10^{44}$ \\
\hline
\end{tabular}
\label{smalltab}}
\end{table}

In Fig.~\ref{red} 
we plot the  i-band magnitudes as a function of redshift. The magnitude distribution is typical  for a flux-limited sample.

\subsection{Comparison with other catalogues}

 It is now interesting to cross-match the Te-REX sources with other catalogues of blazars (or candidates) that recently appeared in the literature (see Tab.~\ref{vis}). In particular, we consider the last preliminary catalogue of gamma-rays sources discovered by Fermi-LAT (FL8Y\footnote{https://fermi.gsfc.nasa.gov/ssc/data/access/lat/8yr\_catalog/4FGL\_Catalog\_v4.pdf}), the 2WHSP catalogue of HBL candidates \citep{chang17}, and the 5th edition of the Roma-BZCAT Multifrequency Catalogue of Blazars (5BZCAT, \citealt{massaro09,massaro15}).
 
The  preliminary 8-year Fermi-LAT source list catalogue (FL8Y) contains
all the sources detected by Fermi LAT  above 4-sigma significance in the first 8 years of observations.
 We cross-checked the two catalogues and we found that 
 out of the 46 Te-REX, only 9 (all classified as BL Lacs) are present in the FL8Y catalogue (less than 20\% of the entire Te-REX sample). 
 
The 2WHSP is a large compilation of HBL candidates obtained by cross-matching the AllWISE whole sky infrared catalog \citep{cutri13} with a number of  radio and X-ray surveys. The selection has been refined  adopting criteria based on 
 IR colours, SED properties, and on the
 visual inspection of multiwavelength SEDs, to ensure that the synchrotron peak is above 10$^{15}$ Hz. Moreover,
 the 2WHSP includes also blazars emitting at VHE.
 The selection in the IR colours is a technique very effective in selecting powerful
 blazars that populate a delimited region of the WISE colour-colour diagram,  the so-called "blazar strip" \citep{massaro12}. However, it is less efficient in selecting  
 weak blazars where the IR emission of the star light  of the host galaxy dominates with respect to the non-thermal continuum.
These weak blazars  are largely missed in the 2WHSP catalogue because of  selection criteria, but they are instead a numerous population
in the  Te-REX sample. Indeed, only 1 PEG (out of 21) is present in the 2WHSP catalogue while most of the BL Lacs (20 out of 25) are included. In this sense, the Te-REX sample is more representative with respect to the 2WHSP since it is not biased against the host-galaxy dominated HBL.

The 5BZCat \citep{massaro15} contains blazars discovered in multifrequency surveys through an accurate review of the literature. 
As expected, about 60\% (14 out of 25 objects) of the BL Lacs are in common with the 5BZCat while none of the PEGs is present. 

\section{Predicting of the TeV emission}
\label{results}

The analyses presented in the previous sections have shown that a significant fraction of the HBL present in the Te-REX sample could not been detected (yet) by Fermi-LAT and that many of these should have an optical spectrum heavily diluted by the host galaxy light. This may suggest that the future deep observations with current and  future Cherenkov telescopes may reveal an increasing fraction of BL Lacs that have not been considered so far as potential VHE sources. However, we do not expect that all the selected Te-REX will be actually visible at VHE energies, even if they are HBL. 
For this reason it is important to estimate the likely emission expected at TeV energies (0.1-10 TeV) of all the selected Te-REX and to compare it with the sensitivity of current and future Cherenkov telescopes.

Only a small fraction of our targets has been detected by the Fermi-LAT telescope and, therefore, we estimate the VHE emission using the data available at lower energies, typically at radio and X-ray wavelengths. Since the optical emission is often heavily contaminated by the host-galaxy light, we prefer not to use optical data.

To estimate the VHE flux, we decided to follow a purely empirical approach, not related to a specific theoretical model. In particular, we make use of statistical relations derived from a well-defined sample of HBL from the literature (see below). The intrinsic scatter of these relations is partially due to the
intrinsic sources variability and to the non simultaneous radio and X-ray observations. We will take into account these uncertainties when extrapolating the SEDs in the TeV energy range. 
In summary, we proceed as follows:

\begin{itemize}
  
  \item we estimate the position of the synchrotron peak   ($\nu_{Syn.peak}$) of each Te-REX object. Due to the scarcity of valid photometric data for most of the objects, we estimate the  value of $\nu_{Syn.peak}$ using the relatively narrow relation between $\nu_{Syn.peak}$ and the $\alpha_{RX}$;

  \item we estimate the gamma-ray emission at $\sim$3 GeV using a statistical relation between the gamma-ray flux (Fermi-LAT) and the radio emission;

  \item we estimate the slope  $\Gamma$ of the gamma-ray emission using a relation between Gamma and the value of $\alpha_{RX}$;

  \item we extrapolate the gamma-ray flux (at $\sim$3 GeV)   up to very high energies ($>$TeV), using the derived slope and assuming a cut-off energy. This cut-off energy depends on the synchrotron peak position;

  \item finally, we apply the EBL absorption, depending on z, to obtain an estimate of the flux received at Earth in the 0.1-100 TeV energy band.  If the redshift is unknown,  we  assume a redshift value of z=0.3.

    \end{itemize}
    
\subsection{The statistical relations}
\label{stat}
As anticipated above, our strategy is to estimate the emission at TeV energies using specific statistical
relations observed among the HBL population. 
Since a sensitive blind all-sky survey at TeV energies is not yet available, it is not currently possible to obtain statistical relations to estimate directly the TeV emission starting from a flux at a different wavelength (i.e. with no bias). For this reason, we need an intermediate step, based on gamma-ray  data at lower energies ($<$0.1 TeV) and, then, an extrapolation of the flux at higher energies. At energies below 0.1 TeV we can use the all-sky survey carried out by Fermi-LAT that provides a blind census of the gamma-ray sources.
The existence of statistical relations between, for instance, radio and GeV-TeV gamma-ray emission has been already suggested by some authors (e.g. \citealt{ackermann11,fan16,lico17}). Here we want to quantify this (and other) relationship that can then be used to predict the gamma-ray flux of the Te-REX objects.

\begin{figure}  
\centering{ 
\includegraphics[width=9.0cm]{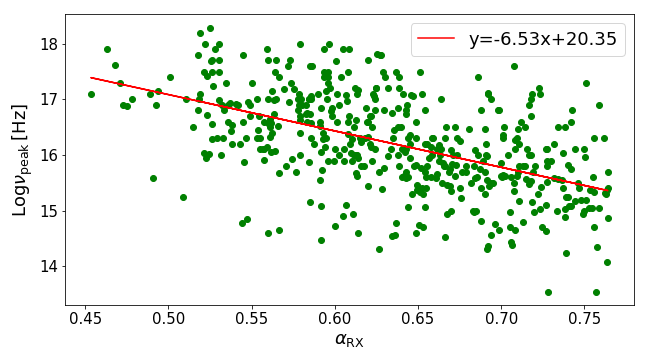} 
\caption{Logarithm of the frequency of the Synchrotron peak for the considered 5BZCAT sub-sample versus the  $\alpha_{\rm RX}$ index.
The red  line is from a  regression analysis.}
\label{alpha}}
\end{figure}

\vskip 0.5cm
\noindent
    {\bf Synchrotron peak frequency vs $\alpha_{RX}$.}
    The first fundamental statistical relation we want to use is the one between the
synchrotron peak and the $\alpha_{RX}$. Such a correlation has been studied several times in the past (e.g. \citealt{wolter98,fossati98,nieppola06})
and it reflects the fact that, depending on the synchrotron peak position, in the X-ray band we observe
different parts of the SED, either the ascending (and relatively weak) part of the Inverse Compton (IC) bump, in low frequency
peaked objects, or the (bright) part of the synchrotron bump, close to the peak, in high frequency peaked objects.
As a consequence, the
X-ray-to-radio flux ratio changes dramatically going from low-frequency to high frequency peaked objects thus
creating a relatively narrow frequency peak vs $\alpha_{RX}$ relation.
In Fig.\ref{alpha} we plot the relationship between the logarithm of the frequency of Synchrotron peak  versus the $\alpha_{\rm RX}$ index using all the BL Lac objects with $\alpha_{RX}<$0.76 present in the  5th edition of the Roma-BZCAT with measured properties.
We perform a linear regression analysis and we find a significant negative relationship between these two quantities (r value=-0.53, correspondent to a probability of 5$\times$10$^{-36}$  that the relation is not significant). Lower $\alpha_{\rm RX}$ indices correspond to higher peak frequencies: the relationship is described by the equation:

\begin{equation}
  Log\nu_{Syn. peak}=-6.53\alpha_{RX}+20.35
  \label{peakarx}
\end{equation}

with a  mean squared error value of 0.55. We apply this relationship to the Te-REX sample, and for each object we derive the frequency of the Synchrotron peak from the $\alpha_{\rm RX}$ value. Since the position of the Inverse Compton and Syncrotron peaks
are related, following \citet{fossati98} we  can roughly estimate the position of the Inverse Compton peak assuming:

\begin{equation}
  Log\,\nu_{\rm Com. peak}\sim Log\,\nu_{\rm Syn. peak}+10
  \label{peakic}
\end{equation}

\begin{figure}  
\centering{ 
\includegraphics[width=9.0cm]{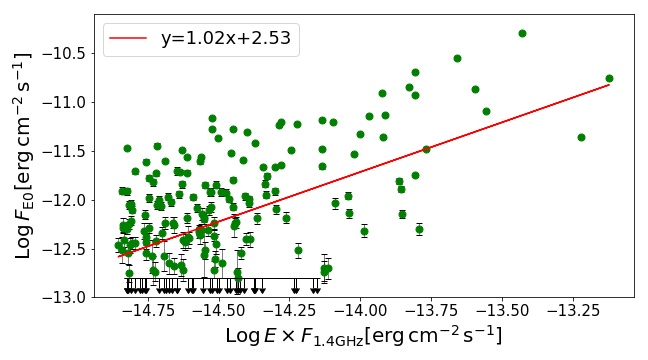} 
\caption{Logarithm of the  Fermi-LAT flux density at the pivot energy versus the radio flux from NVSS or First at 1.4 GHz, 
 above 100 mJy for the considered 5BZCAT sub-sample. 
The red line is from  regression analysis with censored data.}
\label{beta}}
\end{figure}

\vskip 0.5cm
\noindent
    {\bf Gamma-ray flux vs radio flux density.}
    The second fundamental relation is the one between the gamma-ray flux (in the Fermi-LAT energy band)
and the radio flux density. This relation will allow us to estimate the brightness of the sources at high
energies even if the source is not detected by Fermi-LAT. 
To study this relation the choice of the sample is  critical.  Indeed, we cannot use the entire sample of
Fermi-LAT-detected HBL since this is certainly biased versus the brightest gamma-ray sources. Therefore, we
considered all the HBL discovered so far, present in the BZCAT,
independently on their gamma-ray emission, and analysed the
fraction of Fermi-LAT-detected sources as a function of their radio flux density. We found that this fraction
is strongly dependent on the radio flux density, being 60\% at low radio flux densities ($\sim$1mJy) and reaching very
high values ($\geq$90\%) for flux densities above 100 mJy. This is a confirmation that, in HBL, the radio and
the gamma-ray emission are  connected. It is therefore reasonable to estimate the gamma-ray vs radio flux relation using the brightest HBL (above 100 mJy), that have an almost complete gamma-ray detection, and extrapolate it down to low radio flux densities.  This relation can then be used to estimate the expected gamma-ray
flux of all the Te-REX sources, including those not (yet) detected by Fermi-LAT.
In Fig. \ref{beta} we plot the
radio flux density from NVSS or FIRST \citep{white97} at 1.4 GHz versus the
Fermi-LAT flux density at the pivot energy \footnote{The pivot energy
  is the energy at which the error on the differential flux is minimal ($\sim 10^{23.9} Hz$).} $E_0$. As explained above, we consider only the sources with a radio density above 100 mJy in order to have a large fraction of Fermi-LAT detections ($\sim$90\%) thus limiting the
effect of upper limits on the final fit. However, even above a radio flux of 100 mJy,
about 30 sources remains undetected. 
To take into account the presence of  these upper limits (estimated below the sensitivity threshold of Log F=-12.8 erg cm$^{-2}$s$^{-1}$) 
we apply the survival analysis
using the  Astronomy Survival Analysis (ASURV) package \citep{lavalley92} 
implemented in the Space Telescope Science Data Analysis Software (Stsdas) Package \citep{hanisch89}.
Using the method of \citet{buckley79}
for  regressions with  censored data we obtain:

\begin{equation}
  Log F_{E0}=1.02LogF_{1.4GHz}+2.53
  \label{gammaradio}
\end{equation}

Here fluxes are given in erg  cm$^{-2}$ s$^{-1}$.  The mean squared error value is 0.22.

\begin{figure}  
\centering{ 
\includegraphics[width=9.0cm]{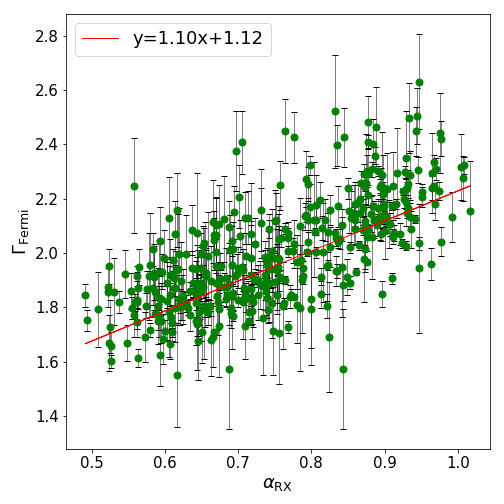} 
\caption{Gamma-ray (Fermi-LAT) slope  versus the  $\alpha_{\rm RX}$ index for the considered 5BZCAT sub-sample.
The red  line is from the  regression analysis.}
\label{gamma}}
\end{figure}

\vskip 0.5cm
\noindent
    {\bf Gamma-ray slope vs $\alpha_{RX}$.}
    The final relation we want to explore is the one
between the slope of the gamma-ray
emission ($\Gamma$) and the $\alpha_{RX}$ value. A relation between
 the slope of the gamma-ray
emission and the logarithm of Synchrotron peak bump has been
already pointed out  by e.g. \citet{abdo10}. Since both  quantities ($\alpha_{RX}$ and $\Gamma$) depend on the value of $\nu_{Syn. peak}$,
this results in a  relatively strict correlation between the gamma-ray slope and the $\alpha_{RX}$ values shown in 
Fig.~\ref{gamma} (correlation value r=-0.68, with a probability of 2$\times$10$^{-57}$ of obtaining this value of r if the two variables were not correlated). The  relation is:

\begin{equation}
  \Gamma=1.10\alpha_{RX}+1.12
  \label{slopearx}
\end{equation}

 The mean squared error value is 0.02.
This relation is very important to predict the slope of
the gamma-ray emission for sources not detected by Fermi-LAT.

\subsection{Estimating the VHE emission}
Thanks to the relations discussed above we can derive, at least in a statistical sense, the expected
VHE emission of the Te-REX sources.
To this end, we model the gamma-ray emission with a power-law and an exponential cut off at the frequency of the Compton
peak:

\begin{equation}
F_{\nu}=F_{\nu_0}\left(\frac{\nu}{\nu_0}\right)^{-\alpha}\times\,e^{-\left(\frac{\nu-\nu_c}{\nu_c}\right)}\,\times\,e^{-\tau_z}
\end{equation}

where
 $F_{\nu_0}$ is the predicted flux density at the Fermi-LAT pivot frequency $\nu_0$, that we estimate from eq.~\ref{gammaradio}, $\alpha=\Gamma-1$ is the
spectral slope, obtained from eq.~\ref{slopearx}, and $\nu_c$ is the frequency of the Compton peak estimated from eq.~\ref{peakarx} and
eq.~\ref{peakic}. 
Finally, the last term depends uniquely on redshift and takes into account the EBL (optical near-IR photons, the primary source of
opacity for $\gamma$-rays)
that interacts with gamma rays.
We convolve the SED with the EBL model of \citet{franceschini17} that derives different curves of opacity for cosmic high energy photons depending on redshifts.

\begin{figure*}  
\centering{ 
\includegraphics[width=5.5cm]{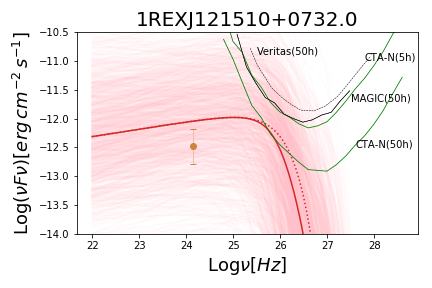}
\includegraphics[width=5.5cm]{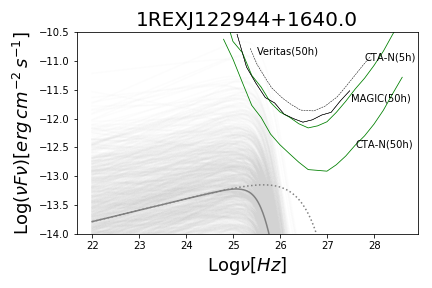}
\includegraphics[width=5.5cm]{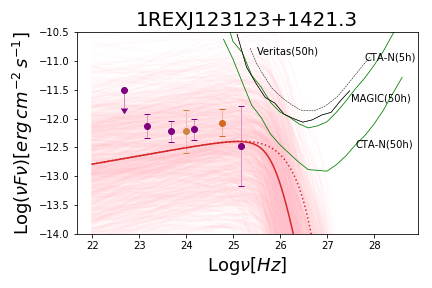}
\includegraphics[width=5.5cm]{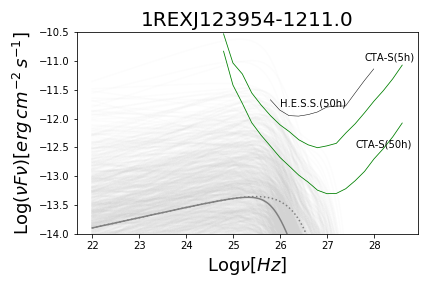}
\includegraphics[width=5.5cm]{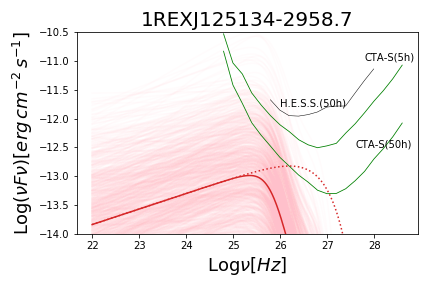}
\includegraphics[width=5.5cm]{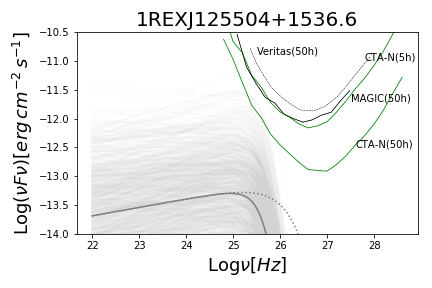}
\includegraphics[width=5.5cm]{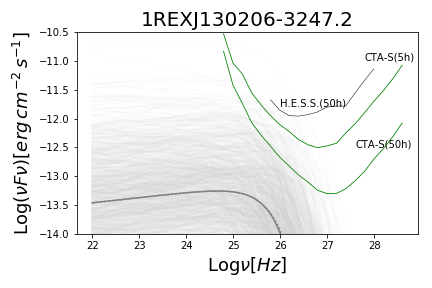}
\includegraphics[width=5.5cm]{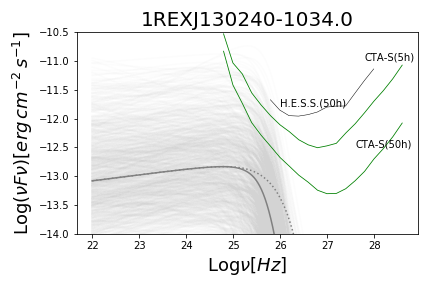}
\includegraphics[width=5.5cm]{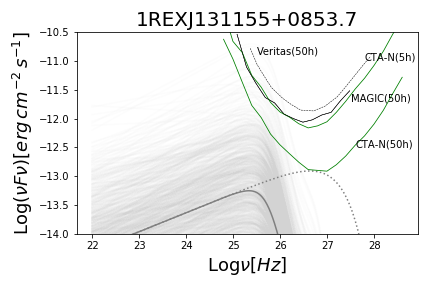}
\includegraphics[width=5.5cm]{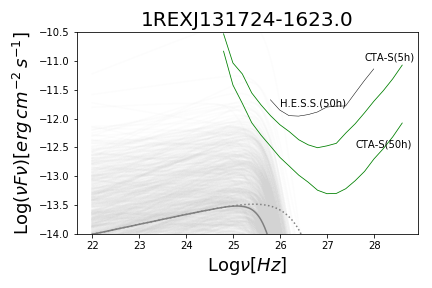}
\includegraphics[width=5.5cm]{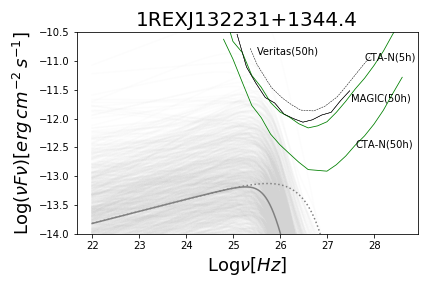}
\includegraphics[width=5.5cm]{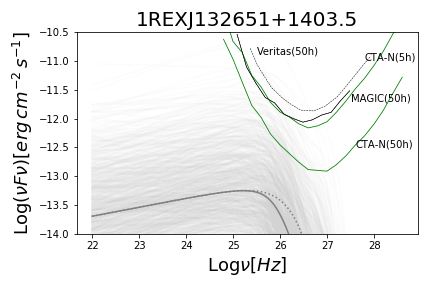}
\includegraphics[width=5.5cm]{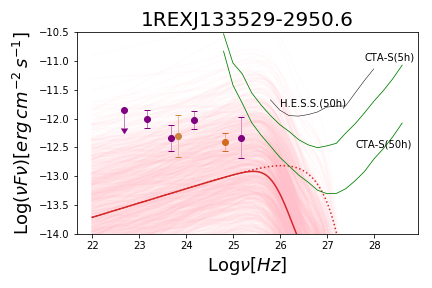}
\includegraphics[width=5.5cm]{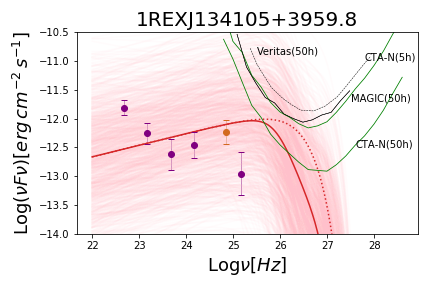}
\includegraphics[width=5.5cm]{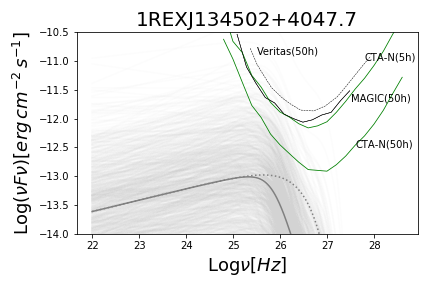}
\includegraphics[width=5.5cm]{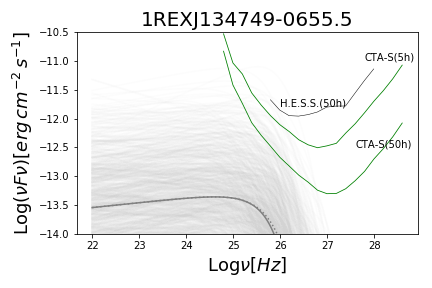}
\includegraphics[width=5.5cm]{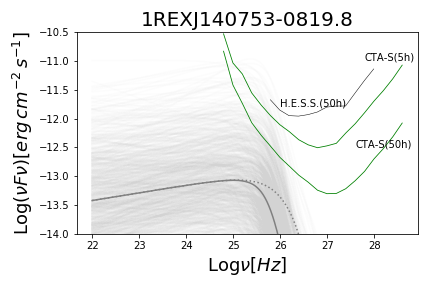}
\includegraphics[width=5.5cm]{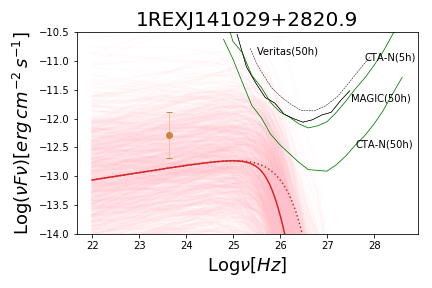}
\caption{Predicted SEDs for the Te-REX sources in the CTA energy observing window.
The SEDs of the best TeV-emitter candidates  are plotted in red (for BL) and in blue (for PEG); the SEDs of non detectable objects are plotted in grey. The dotted and the solid thick lines are the SEDs before and after the EBL model convolution. 
We overplot  1000 likely SEDs (see text for details) and the sensitivity curves  at 5$\sigma$ for 50 and 5 hours for CTA-north or CTA-south dependent on the declination
of the source (solid green lines) and the sensitivity curves  for 50 hours of observation of MAGIC/VERITAS or H.E.S.S.. (black solid for MAGIC or H.E.S.S. and dotted line for VERITAS).
For reference, when available, we superpose the Fermi-LAT observed flux points from the 3FGL (at 200 MeV, 600 MeV, 2 GeV, 6 GeV and 60 GeV) or from the FL8Y.}
\label{allsim}}
\end{figure*} 
 \addtocounter{figure}{-1}
\begin{figure*}
\centering{
\includegraphics[width=5.5cm]{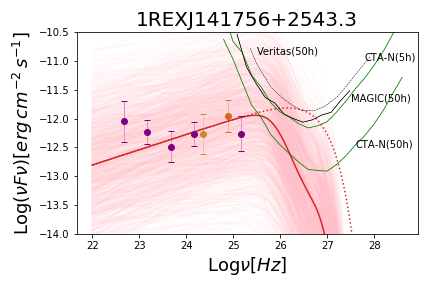}
\includegraphics[width=5.5cm]{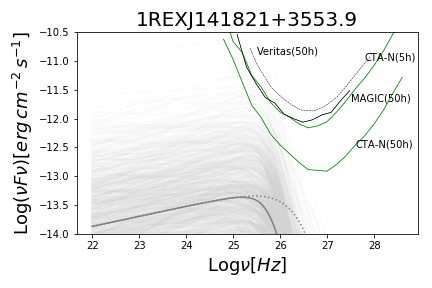}
\includegraphics[width=5.5cm]{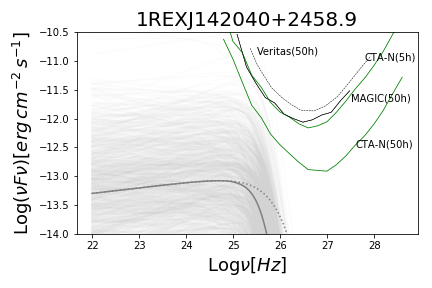}
\includegraphics[width=5.5cm]{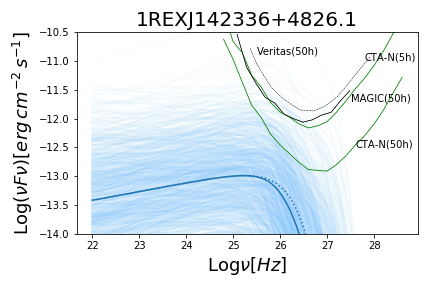}
\includegraphics[width=5.5cm]{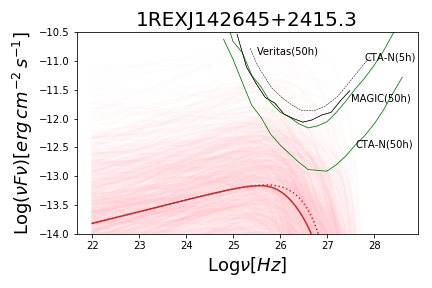}
\includegraphics[width=5.5cm]{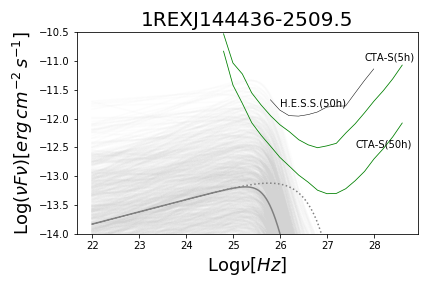}
\includegraphics[width=5.5cm]{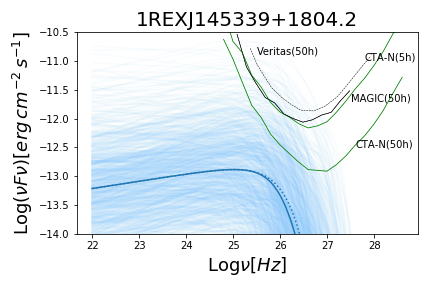}
\includegraphics[width=5.5cm]{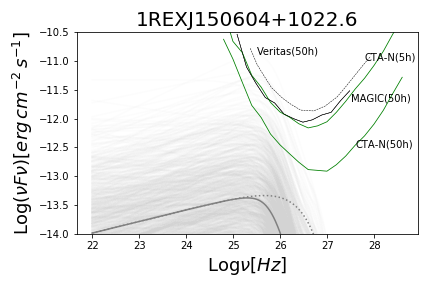}
\includegraphics[width=5.5cm]{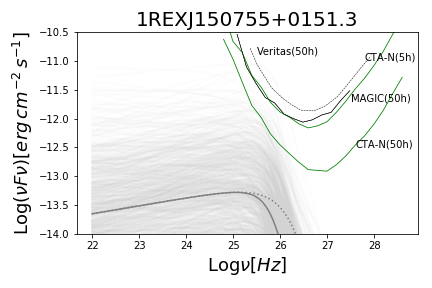}
\includegraphics[width=5.5cm]{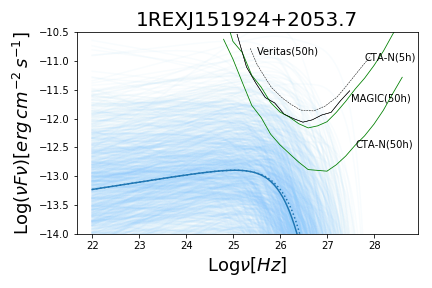}
\includegraphics[width=5.5cm]{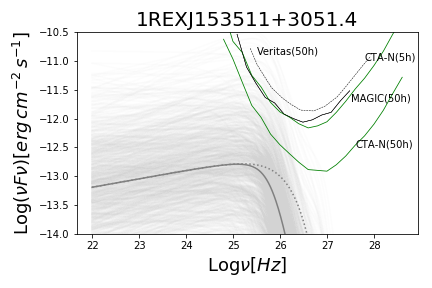}
\includegraphics[width=5.5cm]{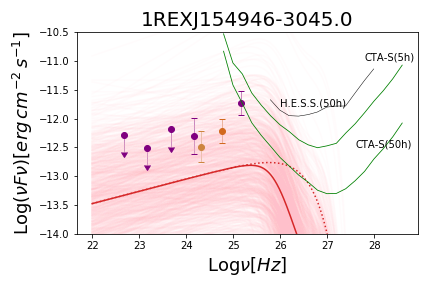}
\includegraphics[width=5.5cm]{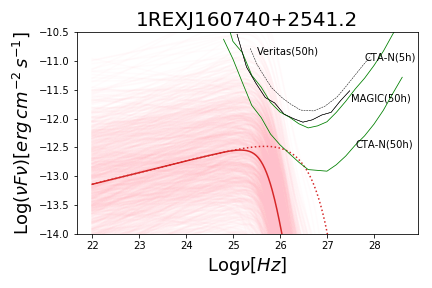}
\includegraphics[width=5.5cm]{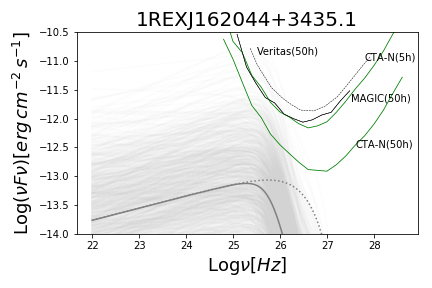}
\includegraphics[width=5.5cm]{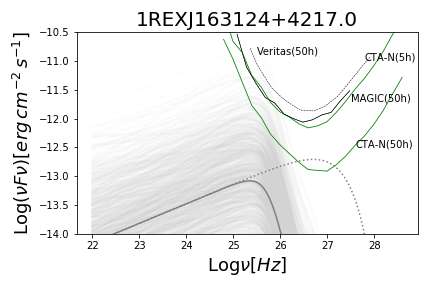}
\includegraphics[width=5.5cm]{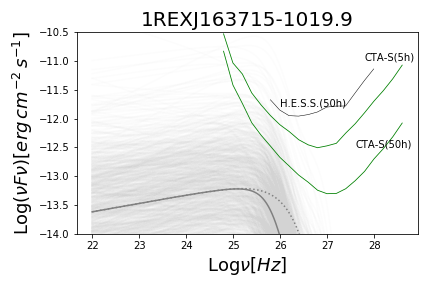}
\includegraphics[width=5.5cm]{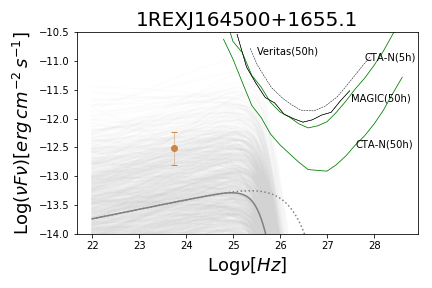}
\includegraphics[width=5.5cm]{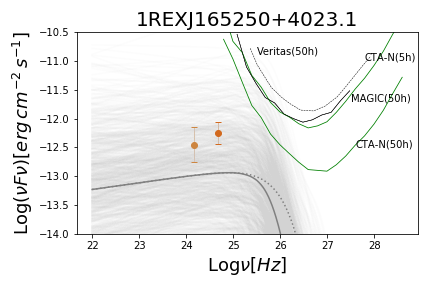}
\caption{continued}}
\end{figure*} 
 \addtocounter{figure}{-1}
\begin{figure*}
\centering{
\includegraphics[width=5.5cm]{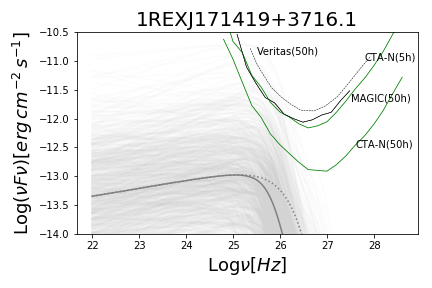}
\includegraphics[width=5.5cm]{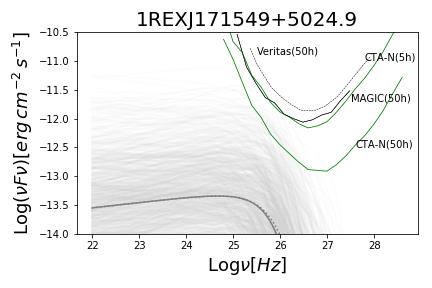}
\includegraphics[width=5.5cm]{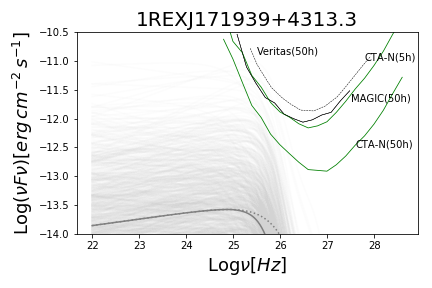}
\includegraphics[width=5.5cm]{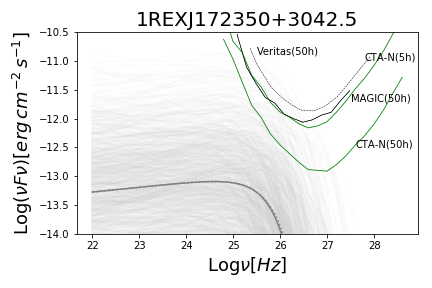}
\includegraphics[width=5.5cm]{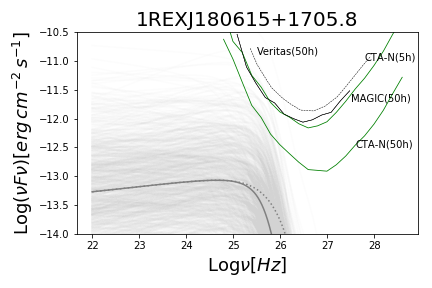}
\includegraphics[width=5.5cm]{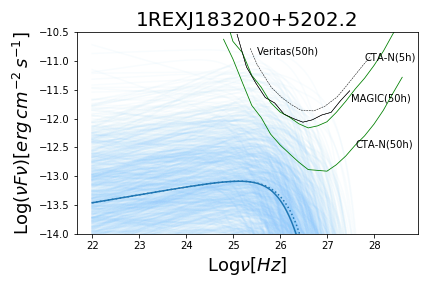}
\includegraphics[width=5.5cm]{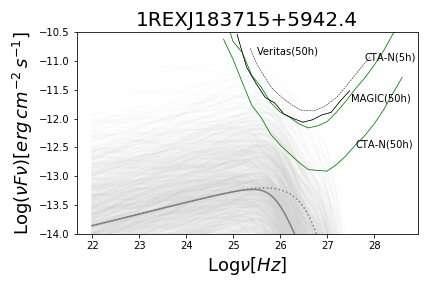}
\includegraphics[width=5.5cm]{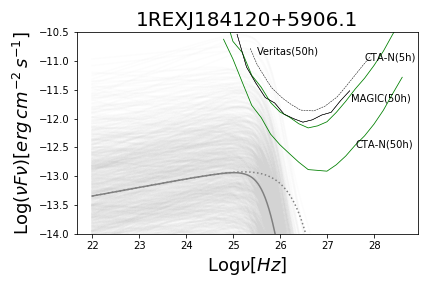}
\includegraphics[width=5.5cm]{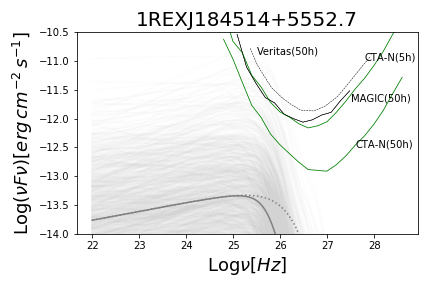}
\includegraphics[width=5.5cm]{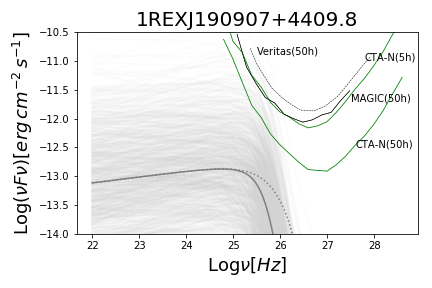}
\caption{continued}
\label{all}}
\end{figure*} 

\begin{table*}
\caption{Cross-match of the Te-REX list with some relevant blazars catalogues from the literature; the percentages are referred  to the sources in common with respect to the number (N) of Te-REX sources.}
\centering{
\begin{tabular}{lccccccc}
      & N & FL8Y & 3FGL& 2WHSP& 5BZCAT&   redshift\\
 \hline

TOTAL  & 46  & 9        & 6        & 21       & 14        & 40 \\
       &     & (20$\%$) & (13$\%$) & (46$\%$) & (30$\%$)  & (90$\%$)\\
            \hline
BL     & 25  & 9        & 6        & 20       & 14        & 19 \\
	   &     & (36$\%$) & (24$\%$) & (80$\%$) & (56$\%$)  & (76$\%$)\\
		\hline
PEG    & 21  & 0        & 0        & 1        &  0        & 21 \\
	   &     & (--)     & (--)     & (5$\%$)  &  (--)     & (100$\%$)\\
		\hline
\end{tabular}
\label{vis}}
\end{table*}

\begin{table*}
\caption{HBL in the Te-REX sample that, according to our estimates, could be detectable by current Cherenkov telescopes (M: MAGIC,
V: VERITAS, H: H.E.S.S. or future (CTA) cherenkov telescopes). The three sources in bold face could be visible even in the shallow CTA survey (5 hours exposure). }
\centering{
\begin{tabular}{lcccccccc}
Name & Inst. & Type & F$_{1.4 GHz}$ & F$_{0.5-2 keV}$ & Fermi-LAT  & BZCAT& 2WHSP\\
\hline
        & & &[mJy] & [e$^{-13}$ergs]& & &\\
 \hline
{\bf 1REXJ121510+0732.0}  &  M/CTA-N    &   BL    &  138    & 25.25   &   Y & Y & Y  \\     
1REXJ123123+1421.3         &  M/CTA-N     &   BL    &  56   & 12.0    &  Y & Y & Y \\	  
1REXJ125134$-$2958.7      &  CTA-S         &   BL    &  10  & 14.54   &    &  Y & Y\\
1REXJ133529$-$2950.6      &  CTA-S         &   BL    &  11  & 12.61   &   Y & Y & Y \\	   
{\bf 1REXJ134105+3959.8}  &  V/M/CTA-N &   BL    &  89      & 51.69   &   Y & Y & Y \\	   
1REXJ141029+2820.9         &  CTA-N         &   BL    &  29 &  4.66   &   Y & Y & Y \\	   
{\bf 1REXJ141756+2543.3}  &  V/M/CTA-N &   BL    &  90      & 136.81  &  Y  & Y & Y \\	   
1REXJ142336+4826.1         &  CTA-N         &   PEG &  12   & 3.43    & 	& &  \\	   
1REXJ142645+2415.3         &  CTA-N         &   BL    &  6  & 4.05    &    & Y & Y 	 \\	   
1REXJ145339+1804.2         &  CTA-N        &   PEG  &  17   & 3.25    & 	& &   \\	   
1REXJ151924+2053.7         &  M/CTA-N    &   PEG  &  16     & 3.21    & 	& &   \\	   
1REXJ154946$-$3045.0      &  H/CTA-S    &   BL      &  16   & 10.59   & 	 Y &   &   \\	  
1REXJ160740+2541.2         &  M/CTA-N    &   BL      &  40  & 18.02   &    & Y  & Y  \\
1REXJ183200+5202.2         &  CTA-N        &   PEG  &  10   & 2.36    &       &    &    \\	 
\hline
\end{tabular}
\label{vis2}}
\end{table*}

\section{Te-REX detectability at VHE}

A detailed simulation to assess the detectability of the Te-REX sources with current or future Cherenkov telescopes is 
beyond the scope of the paper, because the statistical relations used to estimate the VHE have large scatters, and 
so the predictions are  very uncertain for the single object.
Nonetheless, it is useful to compare the predicted VHE fluxes, derived in the previous section, with the sensitivity of some Cherenkov telescopes to understand if they are potentially within the reach of the current (or upcoming) telescopes. In particular, it is of major interest the comparison with the expected sensitivity of CTA which will be the most sensitive telescope at these energies in the next future.


In Fig.~\ref{allsim} we present the unabsorbed  SEDs, built as described in the previous Section (dotted line) and after the convolution with the EBL intensity model (thick lines). 
In order to include the large uncertainties of the scatter relations, we consider 1000  SEDs realisations,
 obtained by combining different $\nu_c$, $F_{\nu_0}$ and $\Gamma$ parameter values, each ones independently and randomly drawn from a Normal  Distribution centred on the best fit parameter values. 
 In the figure we also plot the  sensitivity curve of CTA\footnote{We select the sensitivity curve of CTA North or South \citep{cta19} if,
according to the source declination, the source elevation is largest from the north site (latitude: 28.76 N) or from the south site (south latitude: 24.68 S). We apply the same criterium for selecting the MAGIC/VERITAS or H.E.S.S. sensitivity curve. We note that the targets are preferentially located in the Northern Hemisphere, mainly because the NVSS (and, hence, the REX survey) covers the sky down to $\delta=-40^\circ$ north of declination).} and of three main operating Cherenkov telescopes: MAGIC (Major Atmospheric Gamma Imaging Cherenkov) telescope\footnote{
http://wwwmagic.mppmu.mpg.de/}, VERITAS (The Very Energetic Radiation Imaging Telescope Array System)\footnote{https://veritas.sao.arizona.edu} and H.E.S.S. (High Energy Stereoscopic System)\footnote{https://www.mpi-hd.mpg.de/hfm/HESS/}.
To evaluate whether the Te-REX sources are potentially detectable by current or future Cherenkov telescopes  we apply the following criterium: we consider as good candidates all the sources that have at least 16\% of the SEDs above the instrument sensitivity, meaning 160 times out of 1000 (i.e. located in the upper distribution tail of the possible SED realisation  within the 1$\sigma$ confidence  interval).
All the Te-REX that do not satisfy this condition are probably too faint to have any chances to be detected even with CTA.  Of course, some of these objects could be possibly detected in case of strong flares. 

\begin{figure}  
\centering{ 
\includegraphics[width=6.5cm,angle=-90]{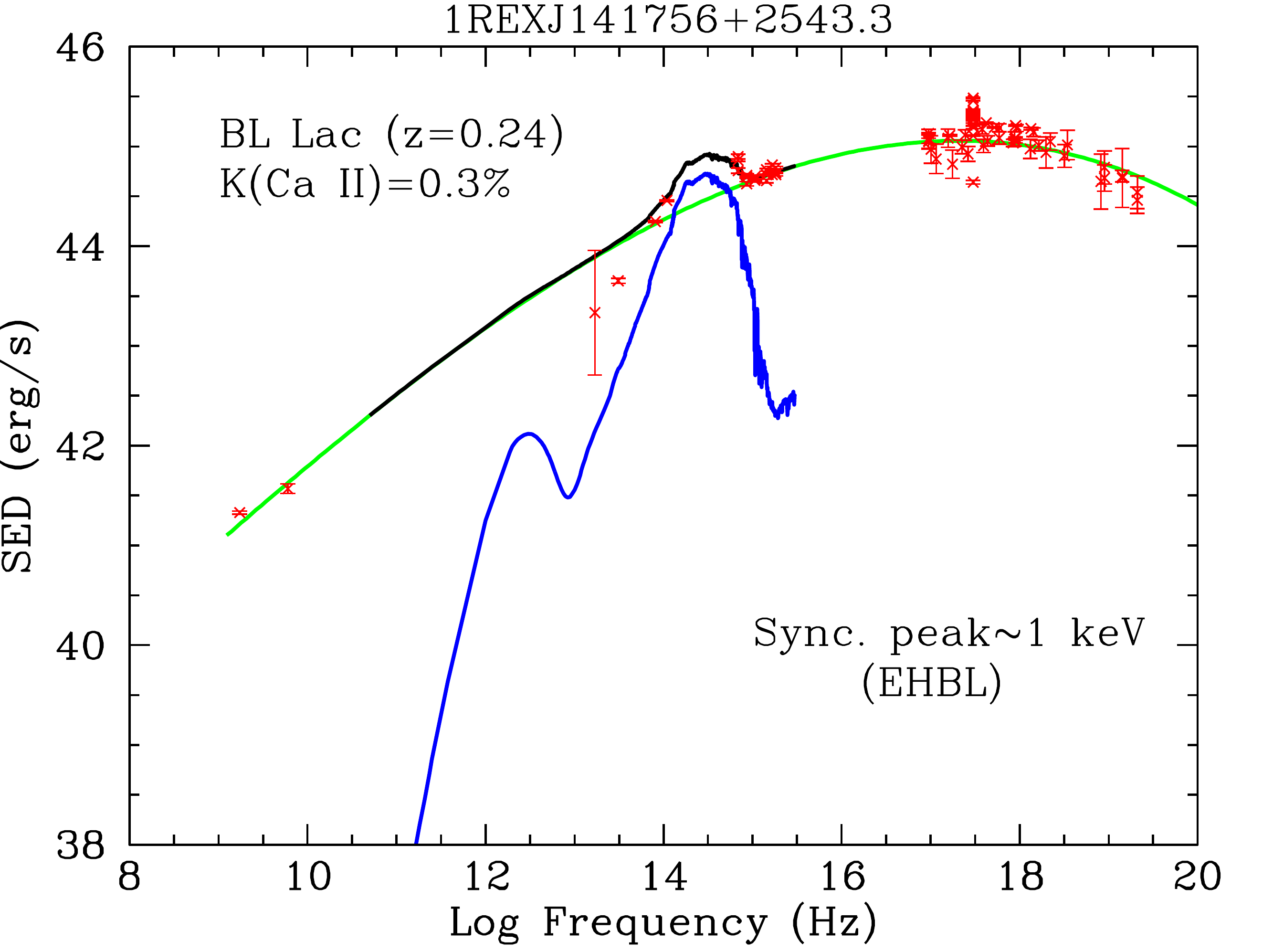}
\caption{SED from radio to X-rays of one Extreme HBL source in the Te-REX sample.}
\label{ehbl}}
\end{figure}

The total number of sources that fulfil the relations, and that we consider as potential candidates for VHE detections, are 14: ten are BL and 4 are PEGs.  In Table \ref{vis2} we report their properties. Furthermore, according  to this analysis we expect that three Te-REX (1REXJ121510+0732.0, 1REXJ134105+3959.8, 1REXJ141756+2543.3) could be visible even with a relatively shallow observation (5h) of CTA or 50h of MAGIC. These are three known HBL sources, also detected by Fermi-LAT and present in the BZCAT and 2WHSP catalogues, with
 a clear BL Lac spectrum (break below 40\%). 1REXJ141756+2543.3 is an extreme HBL (EHBL, \citealt{costamante01}) since the Synchrotron bump peaks at $\sim$1keV (see Fig.~\ref{ehbl}).
 
 Considering deep, pointed CTA observations of 50 hours, 14 targets should be detectable. 
Unlike the three brightest objects discussed above, the 11 additional possible detections have not all already been detected by Fermi-LAT and they are not all {\it classical} BL Lacs from the optical point of view. In particular, only 4 are present in the last version of the Fermi-LAT catalogue (FL8Y), the remaining being still undetected. Even among the objects detected by Fermi-LAT the emission at GeV energies is sometimes very weak as in the case of 1REXJ154946-304. This is a new Fermi-LAT identification that will be discussed in more details in the Appendix \ref{astri}. In addition, almost 1/3 of the 11 Te-REX that should be detected in the deepest observations are classified as PEG
(1REXJ142336+4826.1, 1REXJ145339+1804.2, 1REXJ151924+2053.7, 1REXJ183200+5202.2).

This suggests that, as the observations at VHE will become deeper and deeper, an increasing number of "unusual" BL Lacs could be detected, namely sources with no obvious signature of the non-thermal, nuclear emission in the optical band and very weak, or even absent, emission at Fermi-LAT energies. 

We stress again that, given the  large uncertainties associated to the extrapolation from
statistical correlation, this method will not work for every single source, but could be useful to statistically evaluate the chances of detection of the entire HBL population.

\subsection{Testing the method}
\begin{figure}  
\centering{ 
\includegraphics[width=8.5cm]{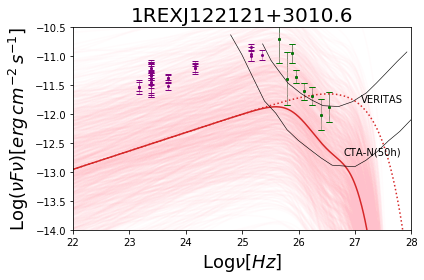}
\caption{
Different SEDs realisations obtained for 
the only REX source detected at VHE (by VERITAS). This source (called in the literature 1ES1218+304) does not belong to the Te-REX sample discussed in this paper because falls outside the area of sky considered here. This object has been observed also by MAGIC but not detected.
We compare the predicted SEDs with Fermi-LAT (in purple) and VERITAS flux points (in green) between $\sim$160 GeV and $\sim$1.8 TeV.}
\label{veritas}}
\end{figure}
It is now interesting to test our method with sources that have been already observed at VHE with the current generation of Cherenkov telescopes. To the best of our knowledge, none of the Te-REX sources considered here have been detected at high energies by HESS, VERITAS or Magic   (mostly because they have not been yet observed):
only 1REXJ134105+3959.8, which is one of the sources that we predict to be detectable with current Cherenkov telescopes,  has been observed by VERITAS for 2.7 hours, providing a flux upper limit in the VERITAS energy band \citet{archambault16}. Given the very low exposure time, the non-detection is not in contrast with our predictions, that were based on 50h of exposure time.

However, considering the entire REX database we have found that one HBL (having a sky position outside the sky area considered in this paper)  is detected by VERITAS (namely 
1REXJ122121+3010.6 or 1ES 1218+304, \citealt{fortin08}). In Fig.~\ref{veritas} we plot for comparison the Fermi-LAT and VERITAS flux points over the 1000 SEDs realisations derived as described in the previous Section. Considering that this source is highly variable and that it is possible that it has been observed by VERITAS in a flaring state, our predictions are quite in agreement with these observations. 
This source is also present in the list of 32 Extreme HBL (EHBL) presented by  \citet{foffano19}. In their sample there is also an other EHBL in common with the Te-REX sources (the previously discussed 1REXJ141756+2543.3), that we predict  should be detectable by current Cherenkov telescopes in 50 hours of observations or by CTA in 5 hours.

\section{Discussion} 

\subsection{Which parameter drives the Te-REX detectability?}
The main parameter that drives the possible detection with CTA, according to our predictions, is the radio flux: all the HBLs with radio flux above 30 mJy are expected to be detected (in 50h by MAGIC or in 5h by CTA); for radio fluxes above 15 mJy about 50\% of the sources are detectable while at the radio flux limit of the survey (5 mJy) only 30\% of the HBL are expected to be detected.

The  $\alpha_{\rm RX}$ plays also an important role since all the possible detections have  $\alpha_{\rm RX}<$0.7 i.e. they are among the most extreme HBL (although not necessarily EHBL, peak$>10^{17}$ Hz) while none of the  HBL with $\alpha_{\rm RX}$ between 0.7 and 0.74 are expected to be detected. However, a very flat value of  $\alpha_{\rm RX}$ does not necessarily imply the detection at VHE: we have objects in which $\alpha_{\rm RX}$ is very flat ($<$0.5,  i.e. good EHBL candidates) that are not expected to be detected due to their low radio flux density. It should be kept in mind that the relation between  $\alpha_{\rm RX}$ and the synchrotron peak has a large dispersion and not all the EHBL  in the sample are necessarily those with the flattest  $\alpha_{\rm RX}$ (See Fig.\ref{alpha}).

Given the importance of the  $\alpha_{\rm RX}$ for the VHE detection we do not expect many additional CTA candidates among the BL Lacs in the REX survey with  $\alpha_{\rm RX}$ steeper than 0.74 i.e. low or intermediate 
peaked BL Lacs (LBL or IBL),  that we have not considered for the selection of the Te-REX sample. Therefore, the Te-REX sample should be considered reasonably complete in terms of possible TeV emitter candidates.

\subsection{The extragalactic sky at VHEs} 

The Te-REX sample covers only a small fraction ($\sim$1.3\%) of the entire sky. 
By re-scaling all the numbers, we expect that about
 800 HBL should be detectable in the entire extragalactic sky (34000 deg$^2$, correspondent to |b$^{\rm{II}}|\ge$10) at VHE in $\sim$50 hours using CTA. As discussed above, about 50\% of these sources are probably not (yet) detected by Fermi-LAT and many of them could have an optical spectrum heavily diluted by the host galaxy light. 

On the contrary, if we consider the detectability with current Cherenkov telescopes (or in a shallow CTA extragalactic survey) we expect to detect preferentially "classical" HBL, with the non-thermal nucleus clearly visible in the optical spectrum or even dominant with respect to the host galaxy. Moreover, we expect that most of these objects (if not all) are already present in the last release of the Fermi-LAT catalogue. 

We predict that about 50 HBL should be detected in the CTA extragalactic survey consisting of 5h of exposure time on 10000 deg$^2$,
 in very good agreement with the numbers published by \citet{hassan17} ($\sim$50 sources in 5h and in 10000 deg$^2$) while it is intermediate (but broadly consistent, considering the large uncertainties) between the expectations discussed by \citet{padovani15} (100-160 BL Lacs expected) and the values estimated by \citet{defranco17} (24 detections). The predictions by \citet{hassan17} are extrapolated by the Fermi-LAT (3FHL) catalogue and, therefore, they consider only the blazars currently detected by Fermi-LAT. In our work we are considering also Fermi-LAT undetected objects but, as discussed above, the expected detections in 5h exposure times (3 objects) are all present in current Fermi-LAT catalogue (FL8Y). This means that current predictions for a (shallow) CTA extragalactic survey based on Fermi-LAT catalogue should be reliable while, for deeper pointing, the current Fermi-LAT catalogues could miss up to 50\% of objects.

In the very close future e-ROSITA  (extended ROentgen Survey with an Imaging Telescope Array, \citealt{merloni12}) will survey the entire sky, reaching in the soft X-ray band (0.5-2 keV) a flux 100 times fainter than the ROSAT All Sky SURVEY (RASS) data
and with an on-axis spatial resolution of $\sim$15$\arcsec$.  A cross-match of the point sources detected by e-ROSITA with radio catalogues as the NVSS or the Sydney University Molonglo Sky Survey (SUMSS)  will 
provide virtually all the expected $\sim$800 HBL that could be detected by CTA.

\section{Summary and conclusions}
We have presented a well-defined and representative sample of HBL selected from the REX survey which was obtained  by cross-matching radio (NVSS) and X-ray data
(ROSAT-PSPC pointed data).

In particular, we selected all the REX with an X-ray-to-radio flux ratio typical of an HBL and falling in a well defined area of sky covering about 560 deg$^2$. We then completed the spectroscopic follow-up of all these sources
and we focused on the 46 sources showing no emission lines in the optical spectrum. We called these source Te-REX, standing for TeV-emitting REX candidates, since HBL are the most promising sources to be detected at TeV energies. 
Out of these 46 sources, 
25 have a clear non-thermal nucleus detected in the optical spectrum as a significant reduction of the Calcium II break ($<$40\%).
In the remaining 21 Te-REX sources, instead, 
we measured a Calcium II break  value above 40\%, i.e.  the non-thermal nucleus is swamped by the stellar light.
We called these 21 sources PEGs (for "passive elliptical galaxies"). We expect that most PEGs should be faint HBL nuclei dominated by the host galaxy light.

We then evaluated the intensity of the VHE emission of all the 46 Te-REX. 
To this end,  we  compare the predicted SEDs with the sensitivity curves of current and future Cherenkov telescopes.

Considering  deep, pointed observations of 50 hours, we found 14 Te-REX (2.5 objects per 100 deg$^2$, corresponding to $\sim$800 objects on the entire extragalactic sky above $\mid b\mid\ge$10$^\circ$) that could be detected by CTA in 50 hours and 7 by current IACTs in 50 hours.  Considering the shallow (about 5h exposure) CTA survey,
we found that 3 Te-REX sources could be  detectable:
these are all "classical" BL Lac with the non-thermal nucleus clearly detected in the optical spectrum, and they are all already included in the last version of the Fermi-LAT catalogue. 

Interestingly,  about half of the 14 sources that could be detectable by CTA in 50h are not yet detected by Fermi-LAT and 1/3 are the spectroscopically classified as PEGs. This suggests that a significant fraction of all the HBL that will be detected at VHE in the near future are yet to be discovered and hidden among apparently normal elliptical galaxies with a relatively bright ($\gtrapprox$ 10 mJy) radio compact core. These objects can be discovered  by means of their X-ray emission which, however, is expected to be below the threshold of the currently available ROSAT All-Sky-Survey X-rays data. 
 Serendipitous and deeper X-ray data currently available from different telescopes (like XMM-Newton or Chandra) are only able to find a small fraction of these objects, due to the limited area of sky covered.  The incoming all-sky X-ray survey that will be carried out by eROSITA will be deep enough to unveil most of these sources on the entire sky in the next few years.

\section*{Acknowledgements}

We acknowledge support by the Italian National Institute of Astrophysics (INAF) through  
the grant INAF CTA$-$SKA
{\it Astri/CTA Data Challenge} (ref. Patrizia Caraveo). We acknowledge also financial contribution from the agreement {\it ASI/INAF NuSTAR n.I/037/12/0}.

This work is based on observations made with the the UH 88$"$ telescope  at Mauna Kea (USA), at UNAM 2.1m at  San Pedro Martir (Mexico), at ESO 3.6m, 2.2m and 1.5m telescopes in 
 La Silla, (Chile) and at the Telescopio Nazionale Galileo (TNG).TNG is operated on the island of La Palma by the Fundaci$\acute{o}$n Galileo Galilei of the INAF (Istituto Nazionale di Astrofisica) at the Spanish Observatorio del Roque de los Muchachos of the Instituto de Astrofisica de Canarias. We use also data from the Sloan Digital Sky Survey and from the Pan-STARRS1 Surveys.

 Funding for the Sloan Digital Sky Survey IV has been provided by the Alfred P. Sloan Foundation, the U.S. Department of Energy Office of Science, and the Participating Institutions. SDSS-IV acknowledges
support and resources from the Center for High-Performance Computing at
the University of Utah. The SDSS web site is www.sdss.org.

The Pan-STARRS1 Surveys (PS1) and the PS1 public science archive have been made possible through contributions by the Institute for Astronomy, the University of Hawaii, the Pan-STARRS Project Office, the Max-Planck Society and its participating institutes, the Max Planck Institute for Astronomy, Heidelberg and the Max Planck Institute for Extraterrestrial Physics, Garching, The Johns Hopkins University, Durham University, the University of Edinburgh, the Queen's University Belfast, the Harvard-Smithsonian Center for Astrophysics, the Las Cumbres Observatory Global Telescope Network Incorporated, the National Central University of Taiwan, the Space Telescope Science Institute, the National Aeronautics and Space Administration under Grant No. NNX08AR22G issued through the Planetary Science Division of the NASA Science Mission Directorate, the National Science Foundation Grant No. AST-1238877, the University of Maryland, Eotvos Lorand University (ELTE), the Los Alamos National Laboratory, and the Gordon and Betty Moore Foundation. 

We gratefully acknowledge financial support from the agencies and organisations listed here: http://www.cta-observatory.org/consortium acknowledgments. This paper went through internal review by the CTA Consortium and we acknowledge the two referees, Tarek Hassan and Reshmi Mukherjee, for constructive and helpful comments and the anonymous referee for his/her careful report.







\appendix
\label{app}
\section{1REXJ154946-3045.0}

\begin{figure}  
\centering{ 
\includegraphics[width=7cm]{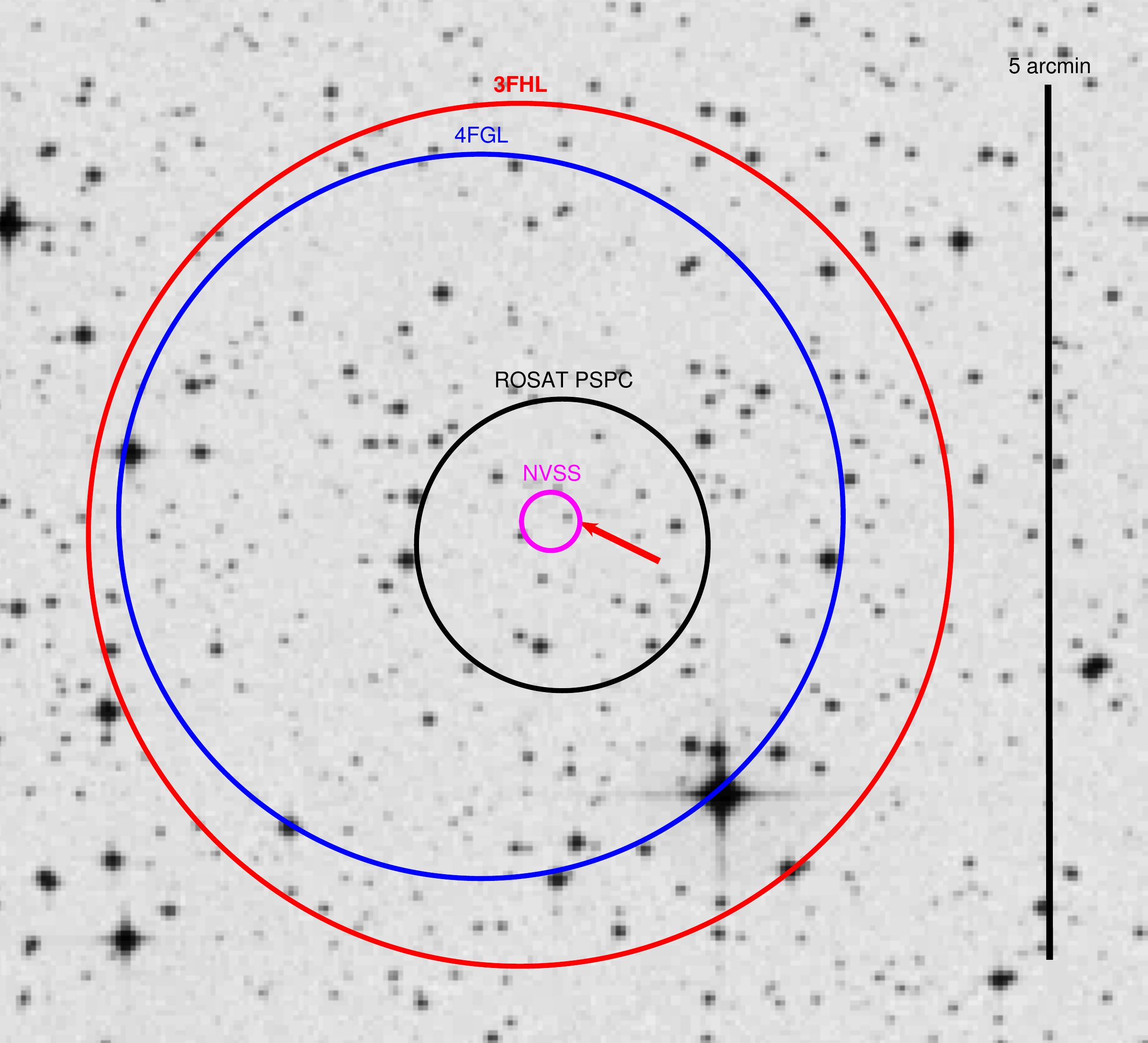} 
\label{astri}
\caption{Finding chart of 1REXJ154946-3045.0. The red, blue, black and magenta circles show the positional uncertainties of the
Fermi-LAT 3FHL, 4FGL, ROSAT-PSPC and NVSS identification, respectively.}}
\end{figure}

The Te-REX source 1REXJ154946-3045.0
is a good example of a candidate for the detection at TeV energies by CTA and, at the same time, it represents a new   identification of the FERMI source 3FGL J1549.9-3044.  The finding chart in Fig. \ref{astri} shows the radio (NVSS) and X-ray (ROSAT PSPC) error circles that define the source as REX. The accurate radio position pin-points the optical counterpart that has been then observed at ESO 3.6m and identified as a BL Lac  with a tentative redshift of z=0.25 (see the optical spectrum in Appendix \ref{spectraall}). The finding chart reports also the position of the 3FHL and 4FGL source which is fully consistent with the REX object making the association certain. 

 In a forthcoming paper, we will evaluate the possible detectability of this interesting source at energies above 10 TeV, with the ASTRI mini-array (Saturni et al., in preparation).

\section{Optical spectra of the Te-REX candidates and main details.}

 We present here the optical spectra of all the Te-REX sources for which an optical spectrum is available (either observed by us or from other repositories, like the SDSS) 
for all the objects (but two, 1REXJ133529-2950.6
and 1REXJ184120+5906.1, for which  the electronic spectrum is not available) classified as BL Lacs of PEGs. 
In the left panel we show the spectrum in unit of erg cm$^{-2}$ s$^{-1}$ \AA$^{-1}$ versus the observed wavelength in \AA\, units. In the right panel
we show a zoom into the Ca~II break region reporting the flux in units of mJy.

The main 
properties for each source are summarised in Table B1. In particular, for each Te-REX source, we provide the sky coordinates (RA and DEC) of the optical counterpart and the name of the counterpart in the NVSS catalogue.

NOTES on individual objects:\\
1REXJ122944+1640.0 uncertain z but similar value in SDSS;\\
1REXJ123123+1421.3  z from \citet{wolter97};\\
1REXJ125134-2958.7  z from \citet{sbarufatti06};\\
1REXJ125504+1536.6 many objects within the radio error box. From FIRST image, the radio emission may come from many distinct objects;
SDSS spectrum of the brightest one at N-E (galaxy at z=0.682). Uncertain identification. It could be a cluster of galaxy;\\
1REXJ131155+0853.7 in SDSS z=0.469;\\
1REXJ132651+1403.5 in SDSS z=0.148;\\
1REXJ133529-2950.6  in \citet{rector00} a different redshift is given (z=0.513);\\
1REXJ133738-2002.5 high density of objects in the optical image: possible cluster;\\
1REXJ133752+2639.1  z based on a single broad emission line (assumed to be MgII$\lambda$2798\AA);\\
1REXJ134502+4047.7 in SDSS z=0.252;\\
1REXJ134749-0655.5  many galaxies in the field: possible cluster of galaxies;\\
1REXJ140414+2846.6 the bright object at S-W is a star;\\
1REXJ141029+2820.9  in SDSS, tentative z=0.52;\\
1REXJ142116+0651.2  QSO at z=1.5 in the X-ray circle but far from the radio position. Possible spurious radio/X-ray association;\\
1REXJ142645+2415.3  a tentative redshift of 0.055 has been proposed by \citet{caccianiga02} and we have assumed this value for the simulation; however, this value is probably too low considering the faint optical magnitude;\\
1REXJ145453+0324.9 in SDSS z=2.302 but it is clearly underestimated;\\
1REXJ150755+0151.3  radio source resolved in FIRST in many components/sources. Possible cluster of galaxies;\\
1REXJ150252-2139.4  two radio sources within the X-ray error box;\\
1REXJ153511+3051.4 nearby galaxy at the same redshift: possible group/cluster of galaxies;\\
1REXJ160347+2048.1 z based on a single broad emission line (assumed to be MgII$\lambda$2798\AA);\\
1REXJ163828+5651.6 NVSS position far from the optical counterpart but FIRST position agrees well with it. Good counterpart;\\
1REXJ165037+5321.1 z based on a single broad emission line (assumed to be MgII$\lambda$2798\AA);\\
1REXJ165250+4023.1 in NED z=0.24 (the origin is unclear). In  \citet{plotkin10}: z$>$0.387;\\
1REXJ171719+4226.9 possible cluster of galaxies from the optical image;\\
1REXJ171322+3256.5 complex radio (NVSS) image. FIRST emission on the counterpart.

\clearpage
\onecolumn
\begin{landscape}
\centering{
\begin{longtable}{ l l l l l l l l l l r r l}
\caption{Column description: (1) source name; (2) Name of the radio counterpart in the NVSS catalogue; (3)  Emission class (BL line AGN, BL or PEG); (4) redshift, we use a question mark symbol to identify uncertain measurements;  (5) apparent magnitude in the i band from the Pan-STARR archive;
(6) {\bf RA and DECof the optical counterpart; }(7) X-ray flux in units of $10^{-13}$ erg cm$^{-2}$ s$^{-1}$; (8) radio flux in mJy;  (9)  radio to X-ray flux density 
ratio $\alpha_{\rm RX}$; (10) Spectroscopical run identifier (see Table \ref{logs}). Otherwise: 
from the Sloan Digital Sky Survey Archive (SDSS);
$^a$ from \citet{ho95};
$^b$ from ZBLLAC, a database of BL Lacs objects, available at the link http://archive.oapd.inaf.it/zbllac/;
$^c$ from the 6dF Galaxy survey \citet{jones04};
$^d$ from the Sedentary survey \citet{giommi05};
$^e$  from The ROSAT International X-Ray/Optical Survey (RIXOS) \citet{mason00} ;
$^f$ classified as BLRG in  \citet{laurent98};
$^g$ published in \citet{caccianiga00};
$^h$ published in \citet{caccianiga02};
$^*$ for this source, the SDSS spectrum is shown even if another is available, for its better quality;
L-M. \citet{laurent98}.\\
} \\
\label{sampleall}\\
\hline
NAME      & NVSS name& Class   &     z  &   i band   &  RA-DEC   &   $F_{\rm x}$                       & $F_{\rm r,ext}$   &   $\alpha_{\rm RX}$  &      REF \\
          &     &    &        &    [mag]   &      &          [10$^{-13}$] & [mJy]        &                     &  & \\
    (1) & (2) & (3) & (4) & (5) & (6) & (7) & (8) &(9) & (10)\\      
\hline

1REXJ103749-2707.2  &NVSSJ103749-270717 & E.L.AGN     &    0.008  & 13.374 & 10:37:49.51 -27:07:16&      5.99 &    16&          0.64 & ESO5/02a   \\
1REXJ121157+0923.8  &NVSSJ121157+092353 & E.L.AGN     &    0.730  & 19.653 & 12:11:57.59 +09:23:57&      1.39 &    12&          0.68 & TNG5/18  \\
1REXJ121510+0732.0  &NVSSJ121510+073205 & BL          &    0.135  & 15.959 & 12:15:10.98 +07:32:05&     25.25 &   138&          0.68 & SDSS$^h$         \\
1REXJ121815+0744.4  &NVSSJ121815+074428 & E.L.AGN     &    0.155  & 17.497 & 12:18:15.46 +07:44:33&      1.03 &     5&          0.67 & MK2/98 $^g$     \\
1REXJ122437+0933.4  &NVSSJ122437+093328 & E.L.G       &    0.520  & 18.177 & 12:24:37.32 +09:33:29&      3.61 &    11&          0.63 & MK3/01    \\
1REXJ122944+1640.0  &NVSSJ122944+164002 & BL          &    0.682: & 19.792 & 12:29:44.51 +16:40:04&      3.84 &    10&          0.62 & SDSS         \\
1REXJ123123+1421.3  &NVSSJ123123+142123 & BL          &    0.260  & 18.210 & 12:31:23.89 +14:21:25&     12.00 &    56&          0.66 & SDSS$^h$           \\
1REXJ123455+1533.9  &NVSSJ123455+153354 & E.L.AGN     &    0.046  & 14.050 & 12:34:55.90 +15:33:56&     17.31 &     7&          0.54 & SDSS         \\
1REXJ123539+1233.3  &NVSSJ123539+123322 & E.L.G       &    0.001  & 10.981 & 12:35:39.86 +12:33:22&     11.15 &   103&          0.72 & $^a$         \\
1REXJ123954-1211.0  &NVSSJ123954-121100 & PEG         &    0.175  & 16.990 & 12:39:54.48 -12:11:04&      2.00 &     5&          0.64 & TNG5/18  \\
1REXJ124205+1117.4  &NVSSJ124205+111727 & E.L.AGN     &    0.407  & 18.637 & 12:42:05.42 +11:17:28&      2.71 &     6&          0.61 & SDSS        \\
1REXJ125134$-$2958.7&NVSSJ125134-295843 & BL          &    0.382  & 18.447 & 12:51:34.84 -29:58:43&     14.54 &    10&          0.55 & $^b$  $^h$         \\
1REXJ125504+1536.6  &NVSSJ125504+153639 & PEG         &    0.682  & 19.157 & 12:55:05.24 +15:36:38&      1.67 &    10&          0.66 & TNG5/18*  \\
1REXJ130206$-$3247.2&NVSSJ130206-324717 & PEG         &    0.031  & 13.807 & 13:02:07.07 -32:47:14&      0.90 &     8&          0.71 & ESO5/02b   \\
1REXJ130240$-$1034.0&NVSSJ130240-103402 & PEG         &    0.495  & 19.463 & 13:02:40.62 -10:34:00&      2.75 &    29&          0.70 & TNG5/18  \\
1REXJ130652$-$1017.6&NVSSJ130652-101738 & E.L.AGN     &    0.95   & 18.027 & 13:06:52.46 -10:17:40&      1.34 &    12&          0.68 & TNG5/18  \\
1REXJ131155+0853.7  &NVSSJ131155+085342 & BL          &    0.48   & 18.900 & 13:11:55.76 +08:53:41&     17.49 &     5&          0.50 & MK2/99$^h$      \\
1REXJ131724$-$1623.0&NVSSJ131724-162304 & PEG         &    0.54:  & 19.768 & 13:17:24.18 -16:23:09&      1.41 &     5&          0.64 & TNG5/18  \\
1REXJ131804+3349.6  &NVSSJ131804+334936 & E.L.AGN     &    0.034  & 15.356 & 13:18:05.42 +33:49:38&      0.78 &     6&          0.70 & TNG5/18  \\
1REXJ131832$-$1811.2&NVSSJ131832-181115 & E.L.AGN     &    0.472  & 17.545 & 13:18:32.91 -18:11:16&      0.97 &     8&          0.68 & $^c$         \\
1REXJ132231+1344.4  &NVSSJ132231+134426 & BL          &    0.376  & 18.029 & 13:22:31.47 +13:44:30&      4.49 &     8&          0.60 & MK2/98    \\
1REXJ132651+1403.5  &NVSSJ132651+140333 & PEG         &    0.147  & 16.188 & 13:26:51.78 +14:03:34&      2.00 &     7&          0.65 & TNG5/18  \\
1REXJ133151+1116.7  &NVSSJ133151+111645 & E.L.AGN     &    0.091  & 16.250 & 13:31:52.23 +11:16:50&      4.75 &     8&          0.61 & $^e$         \\
1REXJ133529$-$2950.6&NVSSJ133529-295036 & BL          &    0.256: & 18.722 & 13:35:29.75 -29:50:39&     12.61 &    11&          0.57 & $^d$   $^h$        \\
1REXJ133738$-$2002.5&NVSSJ133738-200230 & -           &    -      & 20.680 & 13:37:37.67 -20:02:28&      1.47 &    10&          0.70 & $-$         \\
1REXJ133752+2639.1  &NVSSJ133752+263909 & E.L.AGN     &    0.89:  & 20.711 & 13:37:52.28 +26:39:14&      1.00 &     8&          0.67 & TNG5/18  \\
1REXJ134105+3959.8  &NVSSJ134105+395948 & BL          &    0.171  & 16.605 & 13:41:05.11 +39:59:45&     51.69 &    89&          0.61 & SDSS$^h$           \\
1REXJ134502+4047.7  &NVSSJ134502+404746 & PEG         &    0.250  & 18.660 & 13:45:02.79 +40:47:44&      5.61 &    11&          0.61 & MK2/99    \\
1REXJ134749$-$0655.5&NVSSJ134749-065531 & PEG         &    0.115  & 16.658 & 13:47:49.19 -06:55:29&      0.66 &     7&          0.72 & TNG5/18  \\
1REXJ135144$-$3011.7&NVSSJ135144-301145 & E.L.AGN     &    0.248  & 17.774 & 13:51:45.12 -30:11:45&      5.37 &     7&          0.59 & MK2/99    \\
1REXJ135521+3851.0  &NVSSJ135521+385100 & E.L.AGN     &    0.84   & 18.456 & 13:55:20.98 +38:50:58&      1.35 &     7&          0.64 & TNG5/18  \\
1REXJ140414+2846.6  &NVSSJ140414+284637 & E.L.AGN     &    1.41   & 21.000 & 14:04:14.68 +28:46:37&      0.99 &    12&          0.68 & TNG5/18  \\
1REXJ140753$-$0819.8&NVSSJ140753-081951 & BL          &    0.37:  & 18.330 & 14:07:53.50 -08:19:54&      2.33 &    14&          0.67 & TNG5/18  \\
1REXJ141029+2820.9  &NVSSJ141029+282055 & BL          &    -      & 17.650 & 14:10:29.56 +28:20:56&      4.66 &    29&          0.69 & MK2/98$^h$      \\
1REXJ141756+2543.3  &NVSSJ141756+254323 & BL          &    0.237  & 16.664 & 14:17:56.67 +25:43:26&    136.81 &    90&          0.55 & SDSS$^h$\\
1REXJ141821+3553.9  &NVSSJ141821+355356 & BL          &    0.35   & 18.782 & 14:18:21.77 +35:53:55&      1.96 &     6&          0.64 & TNG5/18  \\
1REXJ142034+0704.9  &NVSSJ142034+070458 & E.L.AGN     &    1.504  & 18.486 & 14:20:34.14 +07:04:51&      1.24 &    11&          0.66 & SDSS         \\
1REXJ142040+2458.9  &NVSSJ142040+245856 & PEG         &    0.657  & 19.425 & 14:20:40.05 +24:58:57&      1.44 &    19&          0.71 & SDSS         \\
1REXJ142116+0651.2  &NVSSJ142116+065117 & -           &    -      & 18.421 & 14:21:15.90 +06:51:18&      1.79 &     8&          0.67 & $-$         \\
1REXJ142336+4826.1  &NVSSJ142336+482609 & PEG         &    0.074  & 14.696 & 14:23:36.58 +48:26:10&      3.43 &    12&          0.65 & MK2/98*    \\
1REXJ142347+2404.6  &NVSSJ142347+240439 & E.L.G       &    0.546  & 18.352 & 14:23:47.87 +24:04:42&      6.62 &     9&          0.58 & MK2/99    \\
1REXJ142645+2415.3  &NVSSJ142645+241523 & BL          &    -      & 18.702 & 14:26:45.52 +24:15:23&      4.05 &     6&          0.61 & MK2/98$^h*$      \\
1REXJ143200+4744.8  &NVSSJ143200+474449 & E.L.AGN     &    0.874  & 19.508 & 14:32:01.55 +47:44:55&      3.05 &    20&          0.66 & MK3/01    \\
1REXJ144244+1007.8  &NVSSJ144244+100753 & E.L.AGN     &    1.007  & 17.794 & 14:42:44.99 +10:07:51&      4.79 &    43&          0.67 & MK2/00    \\
1REXJ144436$-$2509.5&NVSSJ144436-250932 & PEG         &    0.39   & 18.482 & 14:44:36.95 -25:09:32&      4.71 &     8&          0.60 & MK2/00    \\
1REXJ145339+1804.2  &NVSSJ145339+180413 & PEG         &    0.071  & 14.692 & 14:53:39.96 +18:04:13&      3.25 &    17&          0.68 & SDSS         \\
1REXJ145453+0324.9  &NVSSJ145453+032456 & E.L.AGN     &    2.375  & 18.137 & 14:54:53.53 +03:24:57&      0.79 &     7&          0.64 & SDSS         \\
1REXJ150252$-$2139.4&NVSSJ150252-213929 & E.L.AGN     &    0.496  & 17.902 & 15:02:52.99 -21:39:29&      2.80 &    21&          0.68 & ESO5/02a   \\
1REXJ150604+1022.6  &NVSSJ150604+102236 & BL          &    -      & 20.421 & 15:06:04.43 +10:22:34&      2.59 &     5&          0.63 & TNG5/18  \\
1REXJ150755+0151.3  &NVSSJ150755+015123 & PEG         &    0.287  & 18.042 & 15:07:54.89 +01:51:22&      1.53 &     8&          0.67 & SDSS         \\
1REXJ151601+0201.0  &NVSSJ151601+020101 & E.L.G       &    0.105  & 15.729 & 15:16:01.38 +02:01:00&     22.62 &    14&          0.56 & MK2/00    \\
1REXJ151924+2053.7  &NVSSJ151924+205345 & PEG         &    0.041  & 12.872 & 15:19:24.74 +20:53:47&      3.21 &    16&          0.68 & SDSS        \\
1REXJ153511+3051.4  &NVSSJ153511+305126 & BL             &    0.320  & 18.101 & 15:35:11.35 +30:51:28&      5.10 &    24&          0.66 & CA7/99*    \\
1REXJ153809$-$0240.4&NVSSJ153809-024029 & E.L.AGN     &    0.094  & 15.732 & 15:38:09.73 -02:40:26&      2.12 &    26&          0.73 & MK4/02    \\
1REXJ154946$-$3045.0&NVSSJ154946-304501 & BL          &    0.245: & 18.809 & 15:49:46.30 -30:45:01&     10.59 &    16&          0.60 & ESO5/02a   \\
1REXJ155535$-$2312.2&NVSSJ155535-231216 & E.L.AGN     &    0.177  & 16.900 & 15:55:34.87 -23:12:18&      5.67 &     7&          0.59 & MK2/00    \\
1REXJ155911+2752.0  &NVSSJ155911+275205 & E.L.AGN     &    1.73   & 18.665 & 15:59:11.76 +27:52:04&      1.91 &     7&          0.61 & TNG5/18  \\
1REXJ160347+2048.1  &NVSSJ160347+204809 & E.L.AGN     &    0.67:  & 20.368 & 16:03:48.14 +20:48:09&      0.69 &     7&          0.69 & TNG5/18  \\
1REXJ160740+2541.2  &NVSSJ160740+254113 & BL          &    0.532  & 17.709 & 16:07:40.60 +25:41:16&     18.02 &    40&          0.61 & MK3/01    \\
1REXJ162044+3435.1  &NVSSJ162044+343511 & BL          &    0.361  & 18.385 & 16:20:44.36 +34:35:11&      5.19 &     9&          0.61 & MK4/02 *   \\
1REXJ162554+5820.0  &NVSSJ162554+582001 & E.L.AGN     &    1.54   & 17.659 & 16:25:55.27 +58:19:58&      1.02 &     7&          0.64 & TNG5/18  \\
1REXJ162752+5419.2  &NVSSJ162752+541914 & E.L.AGN     &    0.316  & 17.490 & 16:27:52.18 +54:19:13&     11.57 &    23&          0.61 & INT5/99   \\
1REXJ162901+4008.0  &NVSSJ162901+400800 & E.L.AGN     &    0.272  & 18.149 & 16:29:01.31 +40:07:60&     11.22 &     9&          0.57 & $^f$           \\
1REXJ162949+0524.0  &NVSSJ162949+052400 & E.L.AGN     &    0.34   & 19.382 & 16:29:49.70 +05:23:58&      4.38 &    31&          0.70 & TNG09/02  \\
1REXJ163020+3756.9  &NVSSJ163020+375657 & E.L.AGN     &    0.394  & 17.085 & 16:30:20.78 +37:56:56&      3.74 &    23&          0.67 & SDSS         \\
1REXJ163057+3707.6  &NVSSJ163057+370739 & E.L.AGN     &    0.801  & 19.258 & 16:30:58.01 +37:07:33&      0.58 &     5&          0.68 & $^e$         \\
1REXJ163124+4217.0  &NVSSJ163124+421703 & BL          &    0.467: & 18.743 & 16:31:24.71 +42:17:02&     29.07 &     7&          0.49 & SDSS$^h$           \\
1REXJ163715$-$1019.9&NVSSJ163715-101957 & BL          &    -      & 19.286 & 16:37:15.65 -10:19:56&      1.91 &     9&          0.68 & TNG5/18  \\
1REXJ163828+5651.6  &NVSSJ163828+565139 & E.L.AGN     &    0.760  & 19.519 & 16:38:30.65 +56:51:36&      0.84 &     7&          0.68 & TNG5/18  \\
1REXJ164500+1655.1  &NVSSJ164500+165510 & BL          &    0.74:  & 17.729 & 16:44:59.80 +16:55:12&      2.19 &    10&          0.65 & MK4/02    \\
1REXJ165037+5321.1  &NVSSJ165037+532111 & E.L.AGN     &    0.551: & 20.013 & 16:50:37.08 +53:21:14&      3.65 &     9&          0.62 & TNG5/18  \\
1REXJ165250+4023.1  &NVSSJ165250+402309 & BL          &    -      & 18.285 & 16:52:49.92 +40:23:10&      2.54 &    19&          0.70 & MK3/01    \\
1REXJ171013+3343.9  &NVSSJ171013+334359 & E.L.AGN     &    0.208  & 15.804 & 17:10:13.42 +33:44:03&      7.62 &     5&          0.55 & SDSS         \\
1REXJ171322+3256.5  &NVSSJ171322+325633 & E.L.AGN     &    0.101  & 16.122 & 17:13:22.59 +32:56:28&      9.66 &    73&          0.70 & CA7/99    \\
1REXJ171419+3716.1  &NVSSJ171419+371610 & BL          &    -      & 19.711 & 17:14:19.78 +37:16:12&      3.01 &    16&          0.68 & TNG5/18  \\
1REXJ171549+5024.9  &NVSSJ171549+502459 & PEG         &    0.097  & 16.445 & 17:15:48.70 +50:25:01&      0.74 &     7&          0.71 & CA7/99    \\
1REXJ171719+4226.9  &NVSSJ171719+422659 & E.L.G       &    0.183  & 19.048 & 17:17:19.20 +42:26:60&     19.02 &   134&          0.69 & SDSS         \\
1REXJ171912+4239.3  &NVSSJ171912+423919 & E.L.AGN     &    1.04   & 18.252 & 17:19:12.77 +42:39:21&      0.87 &    12&          0.70 & TNG5/18  \\
1REXJ171939+4313.3  &NVSSJ171939+431322 & PEG         &    0.425  & 18.613 & 17:19:39.10 +43:13:26&      0.58 &     5&          0.70 & TNG5/18  \\
1REXJ172116+5033.8  &NVSSJ172116+503349 & E.L.AGN     &    0.67   & 19.224 & 17:21:16.23 +50:33:50&      3.79 &    44&          0.73 & TNG09/02  \\
1REXJ172350+3042.5  &NVSSJ172350+304232 & PEG         &    0.047  & 14.251 & 17:23:50.80 +30:42:34&      1.19 &    12&          0.71 & MK3/99*    \\
1REXJ180615+1705.8  &NVSSJ180615+170552 & PEG         &    0.431  & 19.178 & 18:06:15.60 +17:05:49&      1.35 &    17&          0.71 & TNG5/18  \\
1REXJ183200+5202.2  &NVSSJ183200+520217 & PEG         &    0.046  & 14.972 & 18:32:00.65 +52:02:18&      2.36 &    10&          0.67 & CA7/99    \\
1REXJ183715+5942.4  &NVSSJ183715+594226 & PEG         &    0.163  & 16.476 & 18:37:15.91 +59:42:25&      3.53 &     6&          0.61 & CA7/99    \\
1REXJ184120+5906.1  &NVSSJ184120+590609 & BL          &    0.530  & 18.547 & 18:41:20.30 +59:06:08&      3.85 &    20&          0.66 & L-M.     \\
1REXJ184514+5552.7  &NVSSJ184514+555242 & BL          &    0.36   & 19.162 & 18:45:14.10 +55:52:42&      1.60 &     7&          0.66 & TNG5/18  \\
1REXJ190907+4409.8  &NVSSJ190907+440950 & PEG         &    0.518  & 19.585 & 19:09:07.00 +44:09:47&      2.47 &    27&          0.70 & TNG5/18  \\
\smallskip
\end{longtable}}
\end{landscape}
\clearpage
\twocolumn

\begin{figure*}  
\centering{ 
\includegraphics[width=9cm]{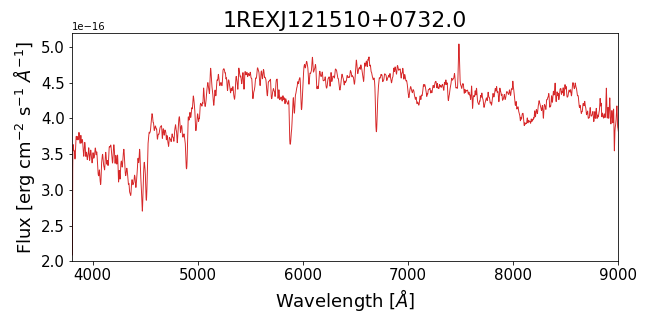} 
\includegraphics[width=4cm]{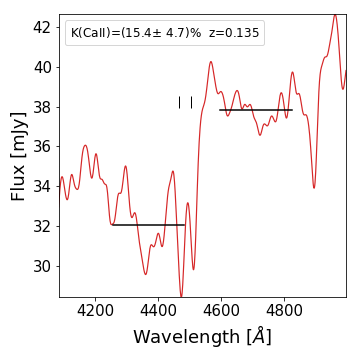} 
\includegraphics[width=9cm]{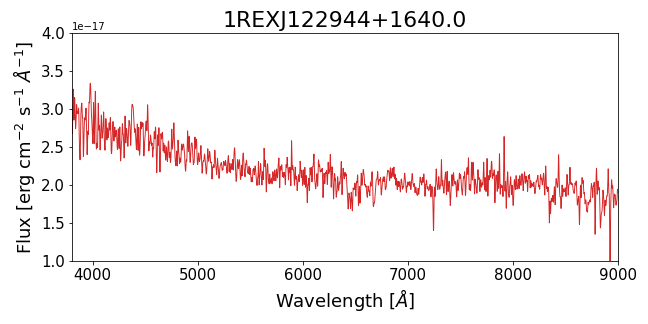} 
\includegraphics[width=4cm]{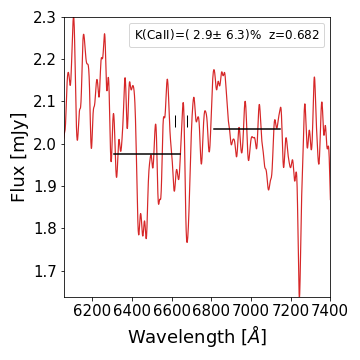} 
\includegraphics[width=9cm]{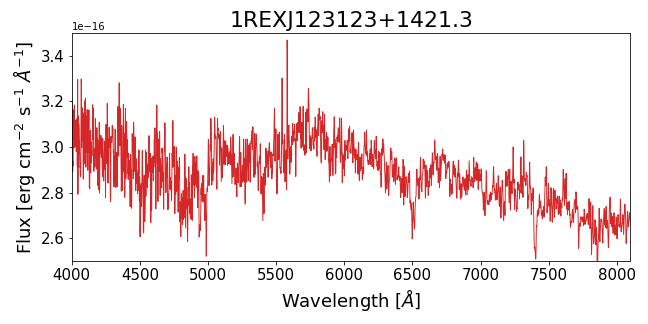} 
\includegraphics[width=4cm]{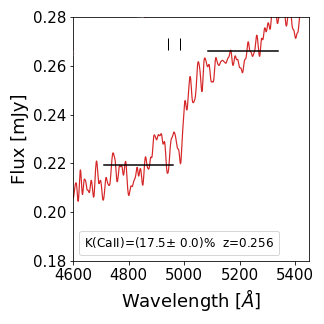} 
\includegraphics[width=9cm]{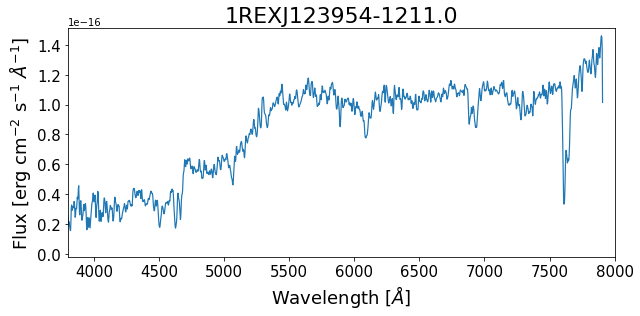} 
\includegraphics[width=4cm]{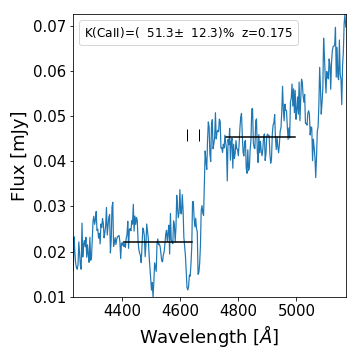} 
\includegraphics[width=9cm]{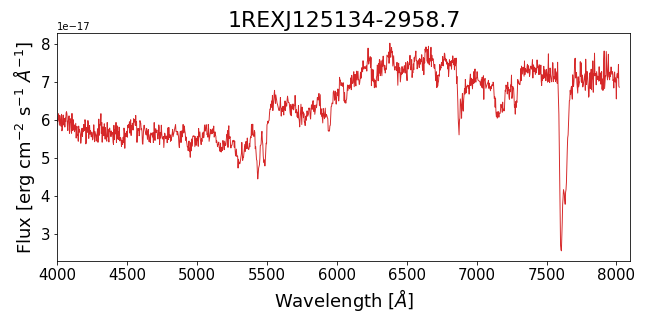} 
\includegraphics[width=4cm]{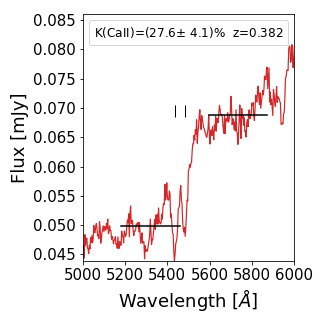} 
\label{spectraall}
\caption{On the left panel we show the optical spectrum for all the Te-REX sources with available spectroscopic data and on the right panel we make a zoom into the Ca~II break region.
We mark the position of the H and K lines of Ca~II and the wavelength interval between 4050\AA\, and 4250\AA\, and between 3750\AA\, and 3950\AA\,
where we measure the Ca~II break. We use the red and blue colours for HBL and PEGs respectively. In the left panel flux  densities are in units of erg cm$^{-2}$ s$^{-1}$ \AA$^{-1}$ and in the right panel in mJy.}}
\end{figure*}  
 \addtocounter{figure}{-1}
\begin{figure*}  
\centering{ 
\includegraphics[width=9cm]{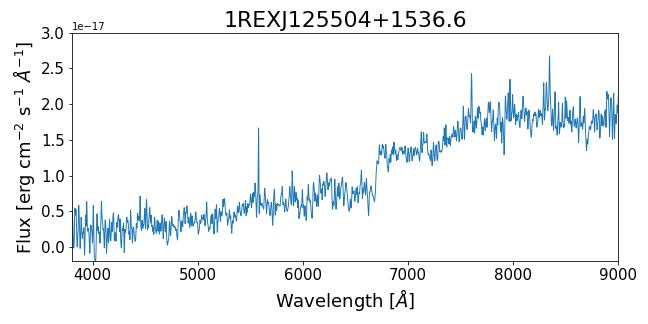} 
\includegraphics[width=4cm]{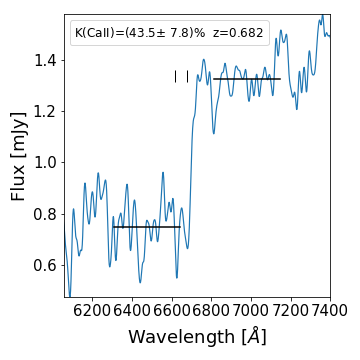} 
\includegraphics[width=9cm]{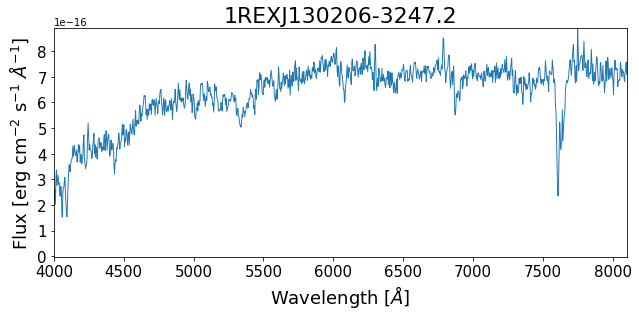} 
\includegraphics[width=4cm]{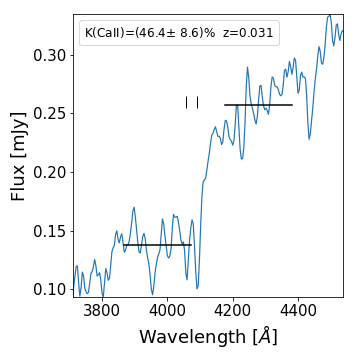} 
\includegraphics[width=9cm]{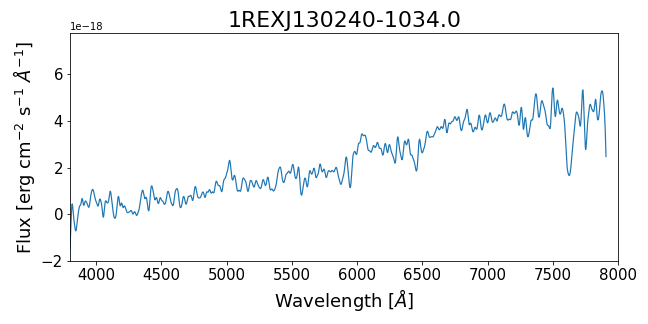} 
\includegraphics[width=4cm]{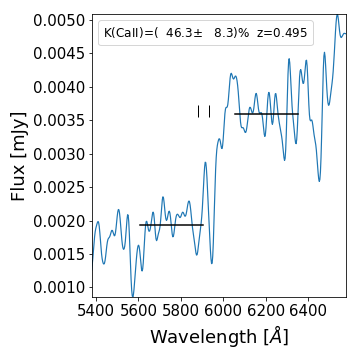} 
\includegraphics[width=9cm]{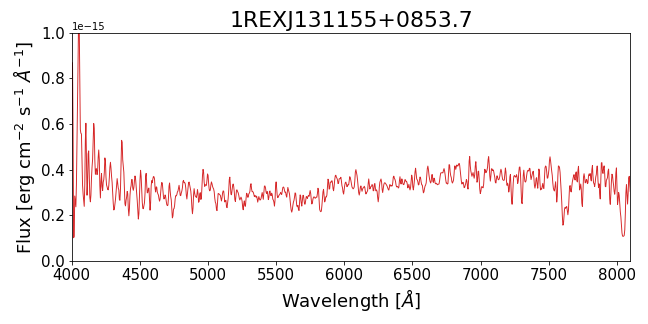} 
\includegraphics[width=4cm]{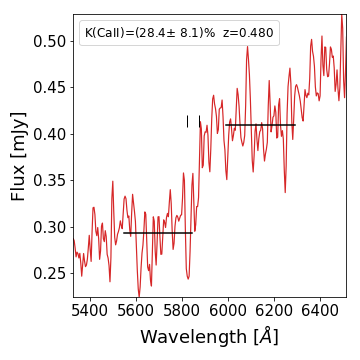} 
\includegraphics[width=9cm]{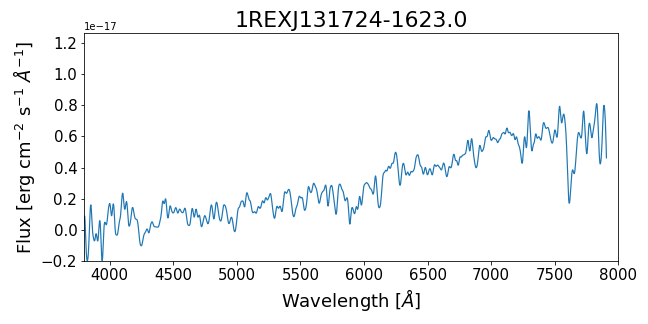} 
\includegraphics[width=4cm]{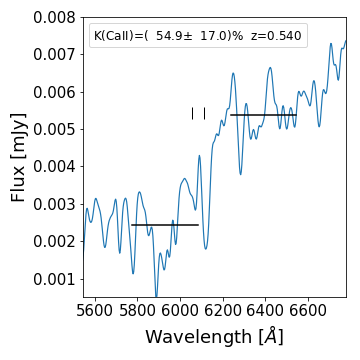} 
\label{nii}
\caption{continued. }}
\end{figure*}   
 \addtocounter{figure}{-1}
\begin{figure*}  
\centering{ 
\includegraphics[width=9cm]{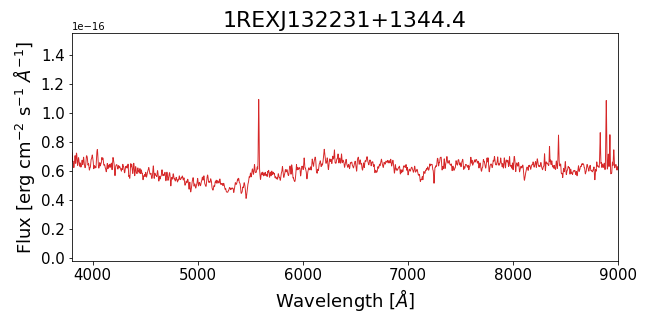} 
\includegraphics[width=4cm]{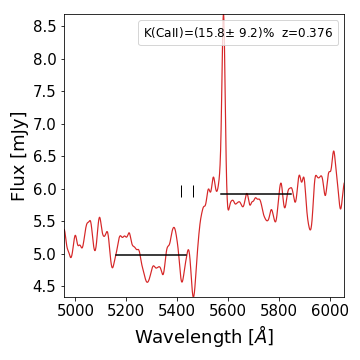} 
\includegraphics[width=9cm]{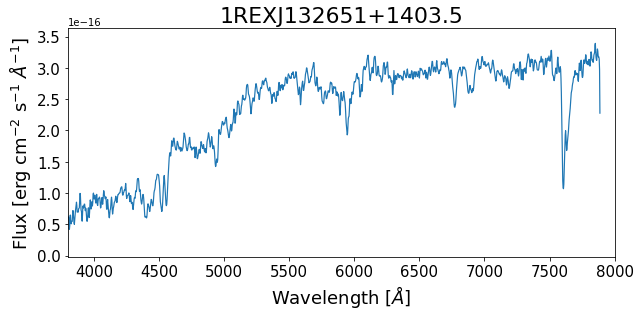} 
\includegraphics[width=4cm]{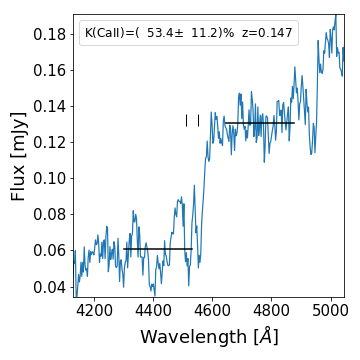} 
\includegraphics[width=9cm]{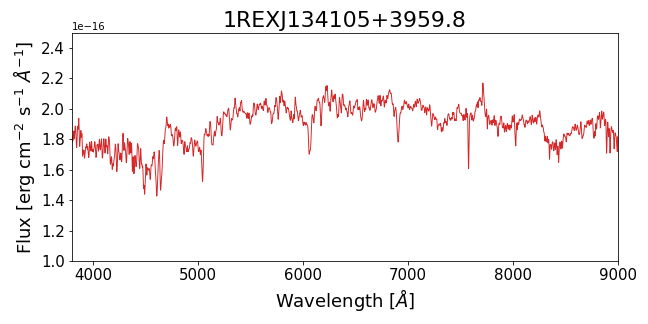} 
\includegraphics[width=4cm]{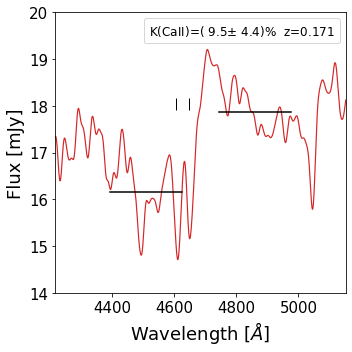} 
\includegraphics[width=9cm]{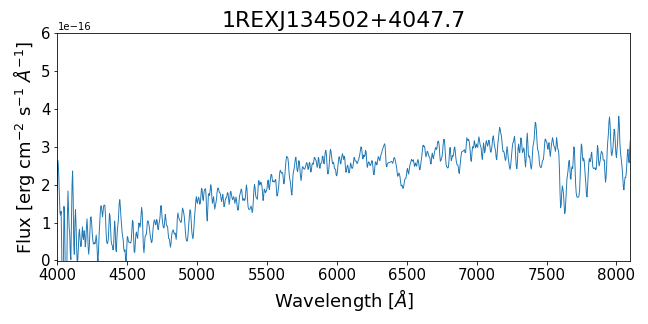} 
\includegraphics[width=4cm]{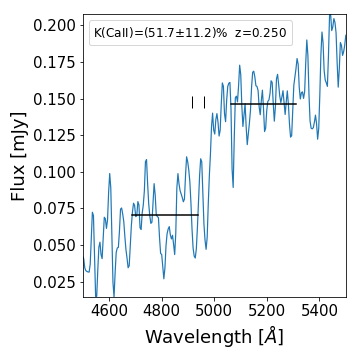} 
\includegraphics[width=9cm]{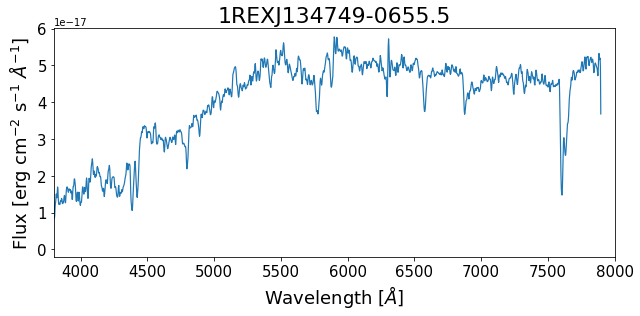} 
\includegraphics[width=4cm]{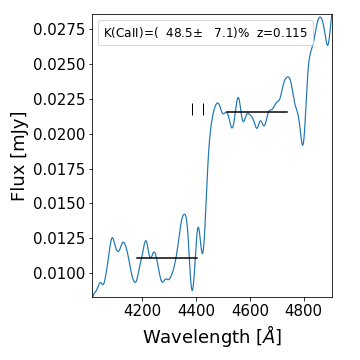} 

\label{nii}
\caption{continued. }}
\end{figure*}   
 \addtocounter{figure}{-1}
\begin{figure*}  
\centering{ 
\includegraphics[width=9cm]{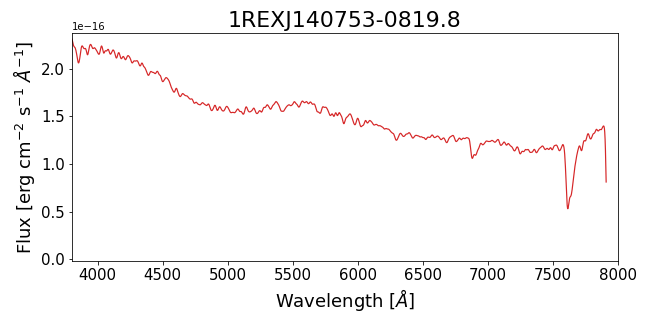} 
\includegraphics[width=4cm]{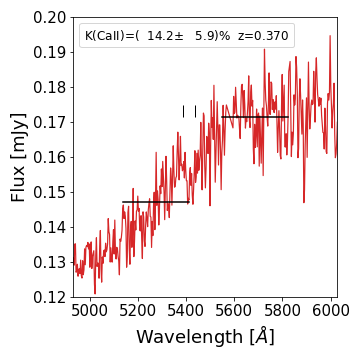} 
\includegraphics[width=9cm]{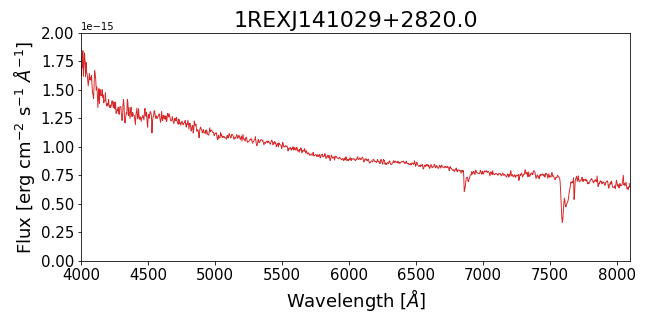} 
\includegraphics[width=4cm]{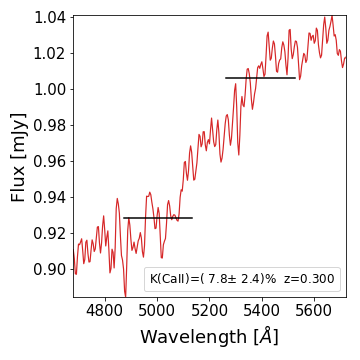} 
\includegraphics[width=9cm]{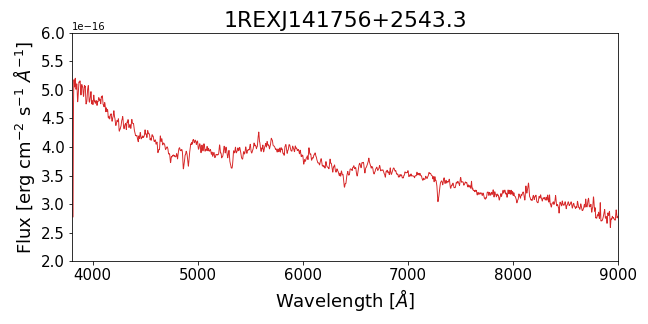} 
\includegraphics[width=4cm]{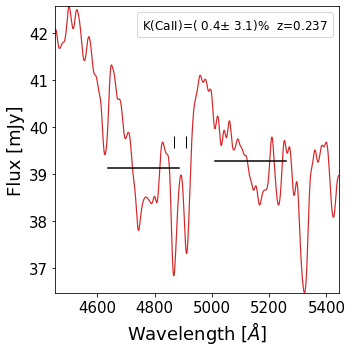} 
\includegraphics[width=9cm]{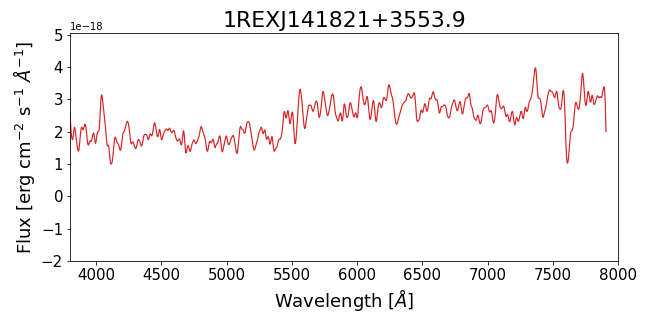} 
\includegraphics[width=4cm]{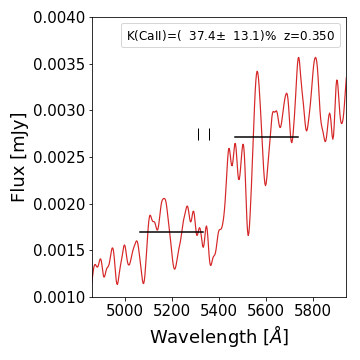} 
\includegraphics[width=9cm]{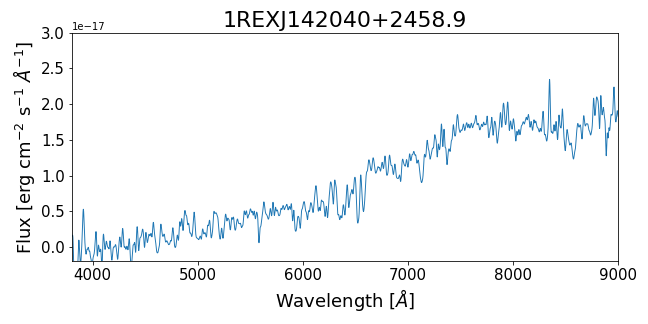} 
\includegraphics[width=4cm]{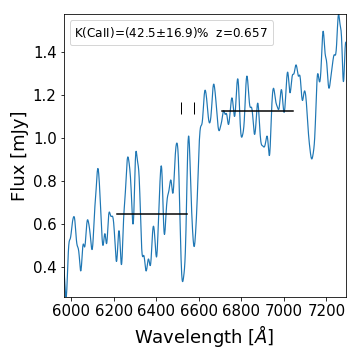} 

\label{nii}
\caption{continued. }}
\end{figure*}   
 \addtocounter{figure}{-1}
\begin{figure*}  
\centering{ 
\includegraphics[width=9cm]{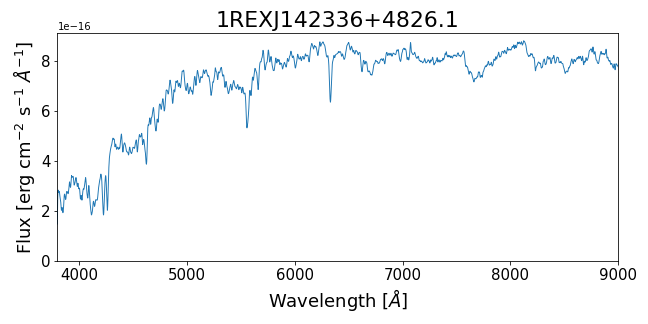} 
\includegraphics[width=4cm]{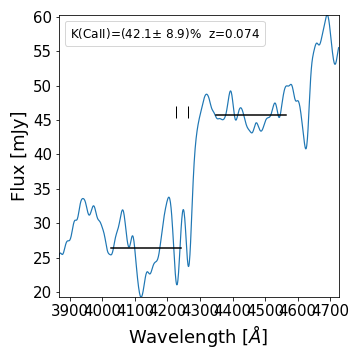} 
\includegraphics[width=9cm]{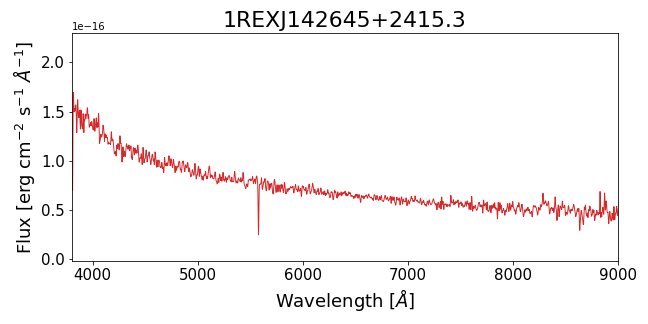} 
\includegraphics[width=4cm]{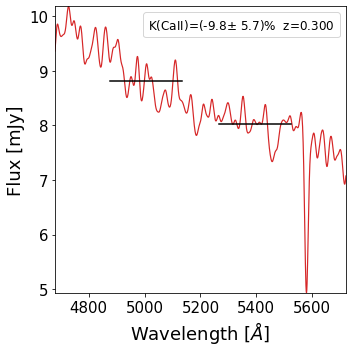} 
\includegraphics[width=9cm]{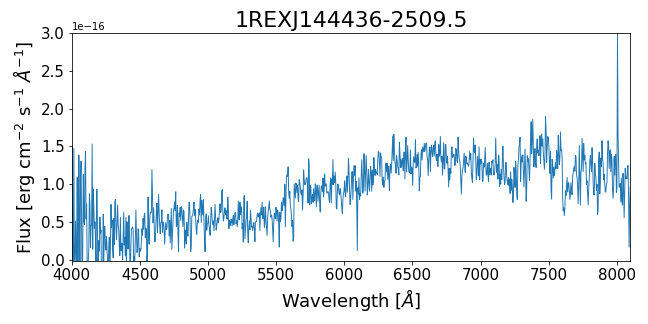} 
\includegraphics[width=4cm]{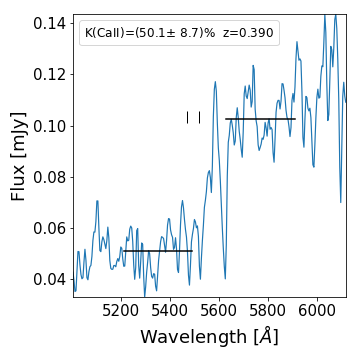} 
\includegraphics[width=9cm]{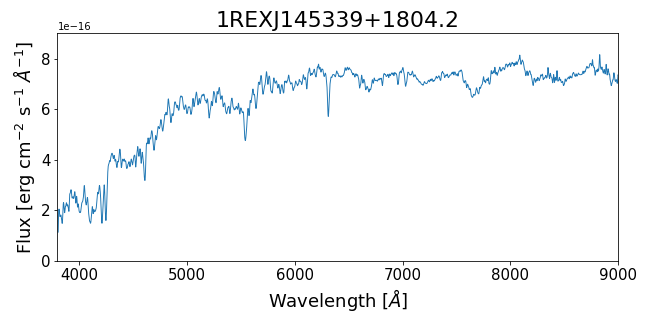} 
\includegraphics[width=4cm]{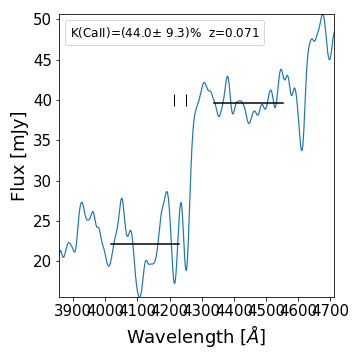} 
\includegraphics[width=9cm]{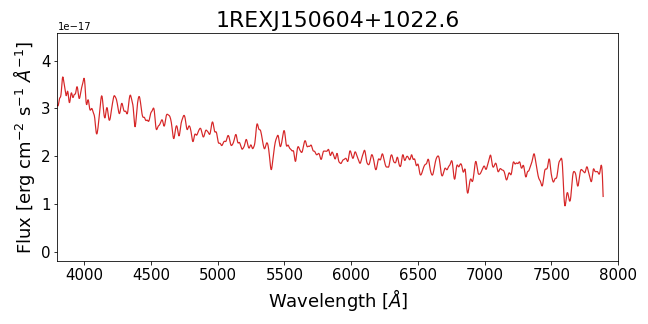} 
\includegraphics[width=4cm]{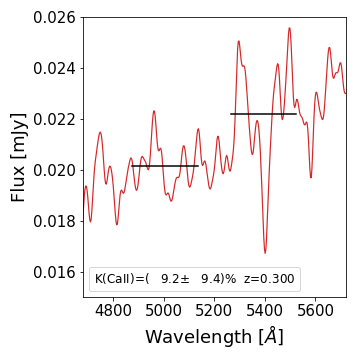} 
\label{nii}
\caption{continued. }}
\end{figure*}  
  \addtocounter{figure}{-1}
\begin{figure*}  
\centering{ 
\includegraphics[width=9cm]{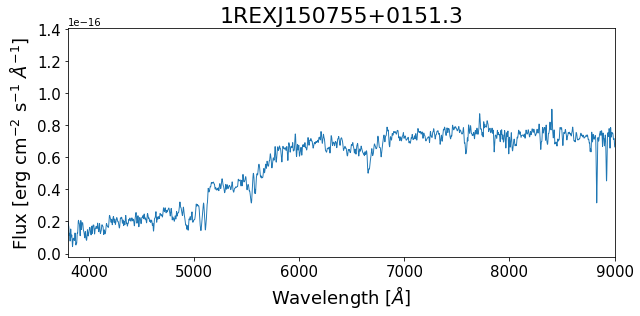} 
\includegraphics[width=4cm]{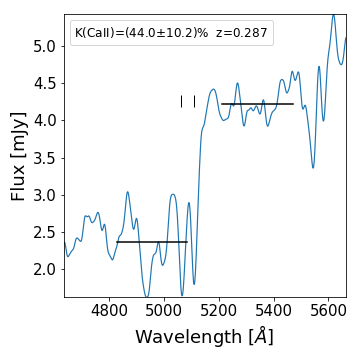} 
\includegraphics[width=9cm]{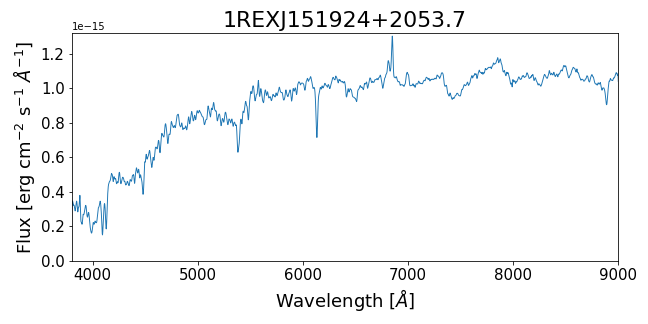} 
\includegraphics[width=4cm]{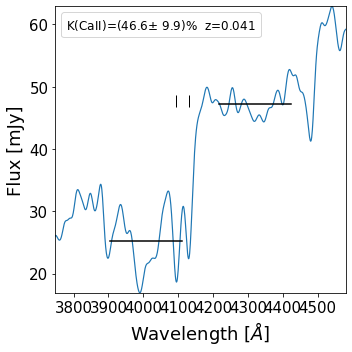} 
\includegraphics[width=9cm]{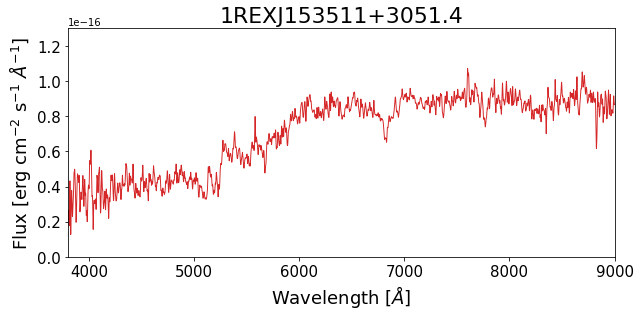} 
\includegraphics[width=4cm]{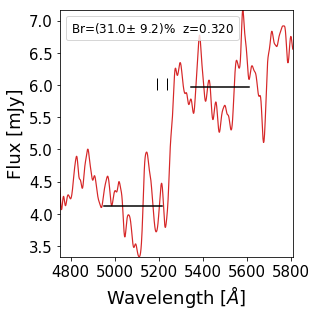} 
\includegraphics[width=9cm]{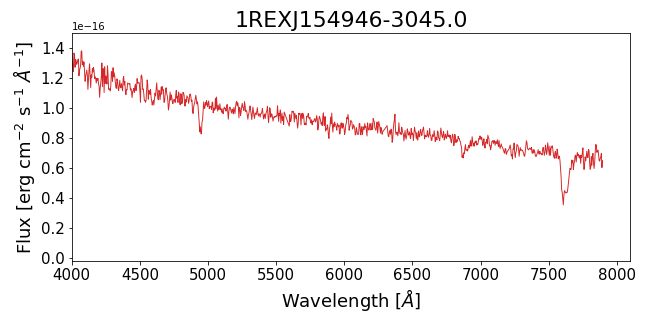} 
\includegraphics[width=4cm]{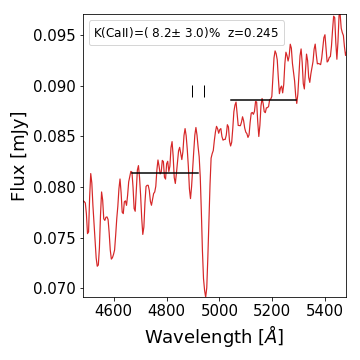} 
\includegraphics[width=9cm]{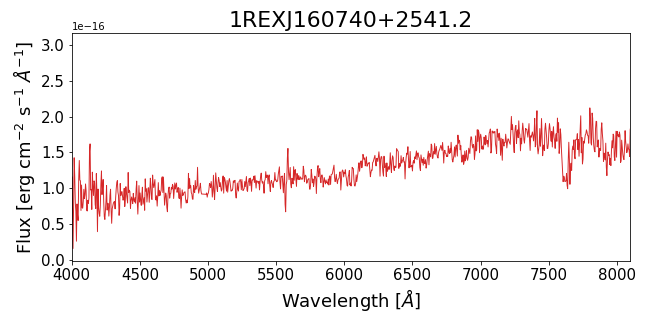} 
\includegraphics[width=4cm]{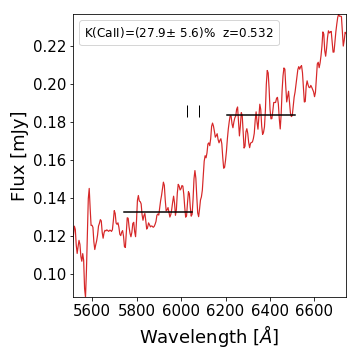} 

\label{nii}
\caption{continued. }}
\end{figure*}   
 \addtocounter{figure}{-1}
\begin{figure*}  
\centering{ 
\includegraphics[width=9cm]{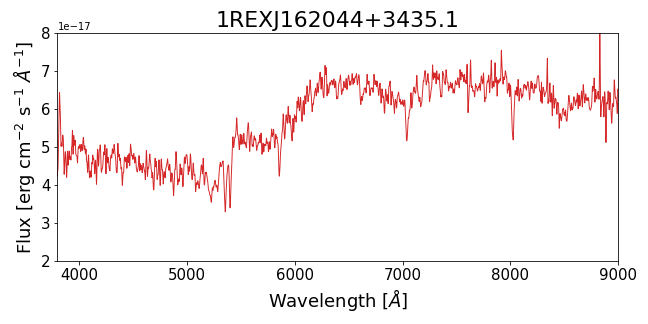} 
\includegraphics[width=4cm]{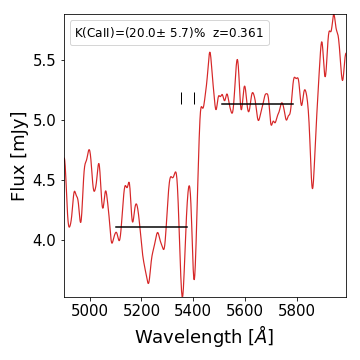} 
\includegraphics[width=9cm]{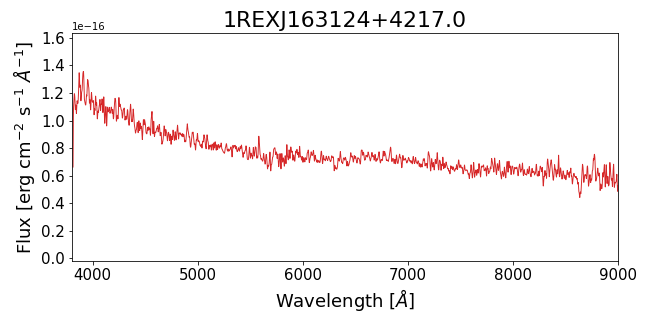} 
\includegraphics[width=4cm]{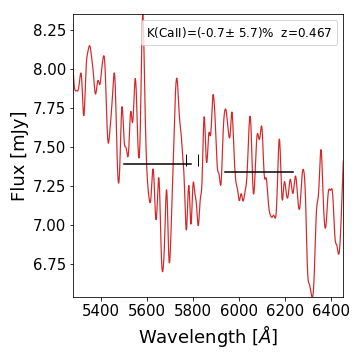} 
\includegraphics[width=9cm]{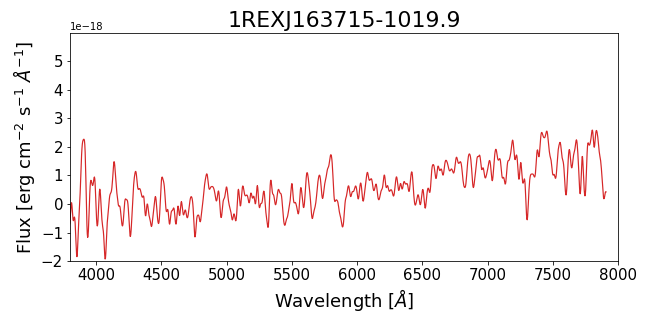} 
\includegraphics[width=4cm]{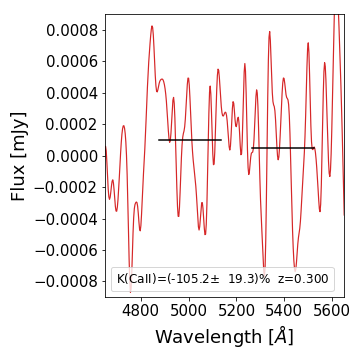} 
\includegraphics[width=9cm]{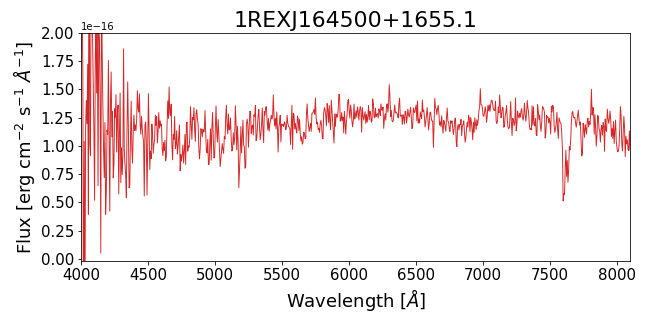} 
\includegraphics[width=4cm]{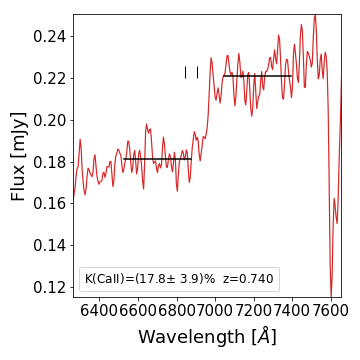} 
\includegraphics[width=9cm]{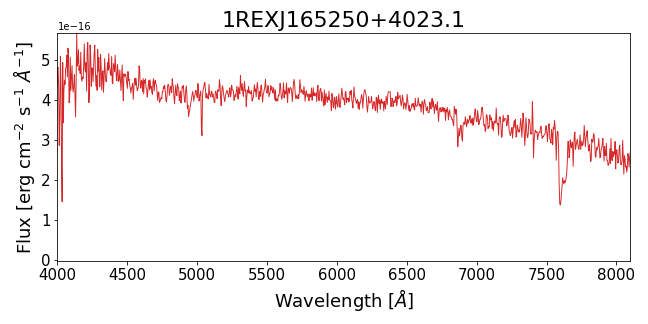} 
\includegraphics[width=4cm]{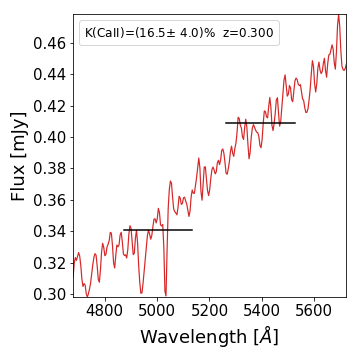} 

\label{nii}
\caption{continued. }}
\end{figure*}   
 \addtocounter{figure}{-1}
\begin{figure*}  
\centering{ 
\includegraphics[width=9cm]{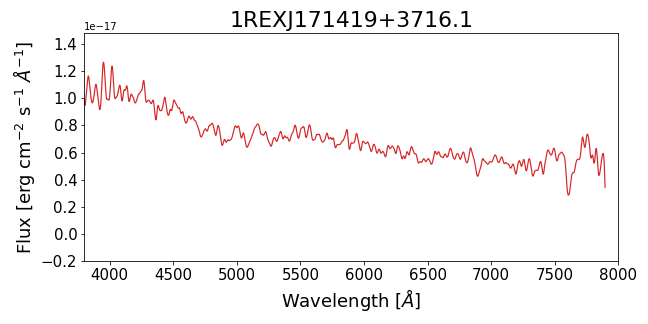} 
\includegraphics[width=4cm]{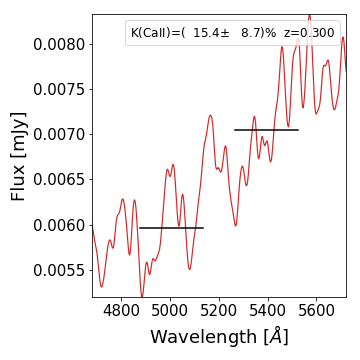} 
\includegraphics[width=9cm]{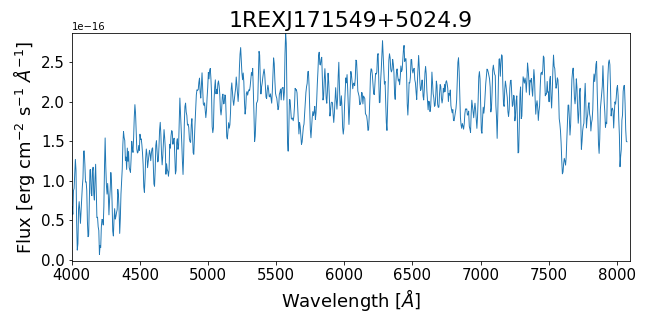} 
\includegraphics[width=4cm]{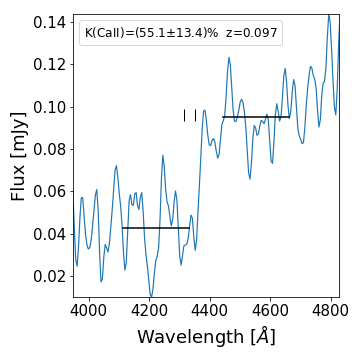} 
\includegraphics[width=9cm]{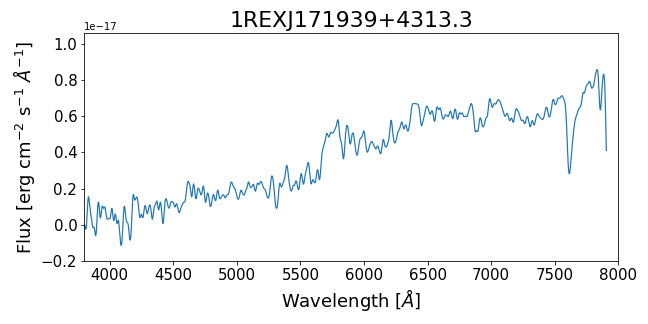} 
\includegraphics[width=4cm]{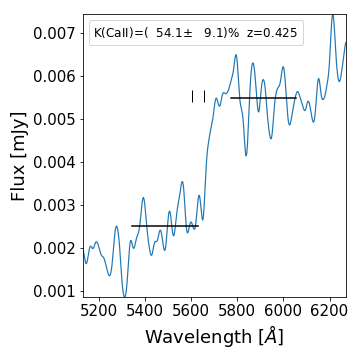} 
\includegraphics[width=9cm]{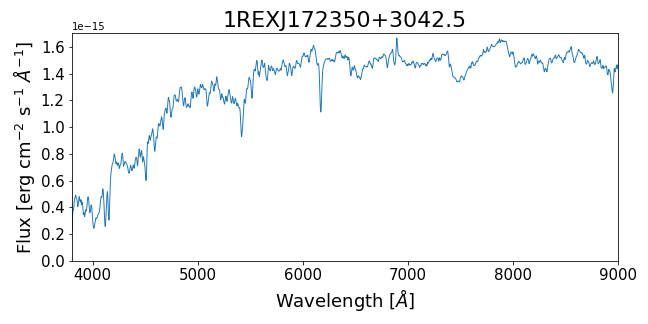} 
\includegraphics[width=4cm]{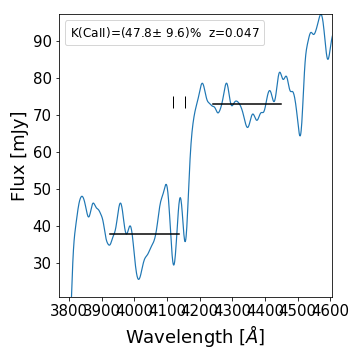} 
\includegraphics[width=9cm]{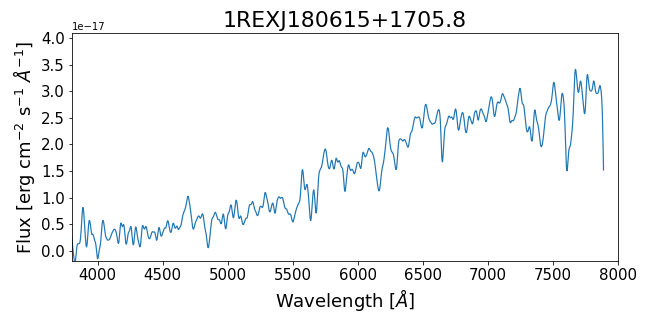} 
\includegraphics[width=4cm]{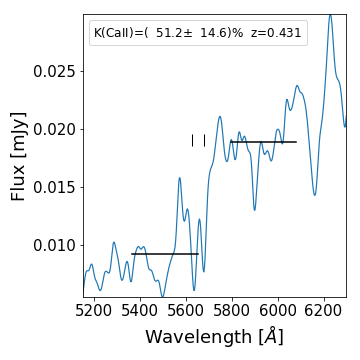} 
\label{nii}
\caption{continued. }}
\end{figure*}   
 \addtocounter{figure}{-1}
\begin{figure*}  
\centering{ 
\includegraphics[width=9cm]{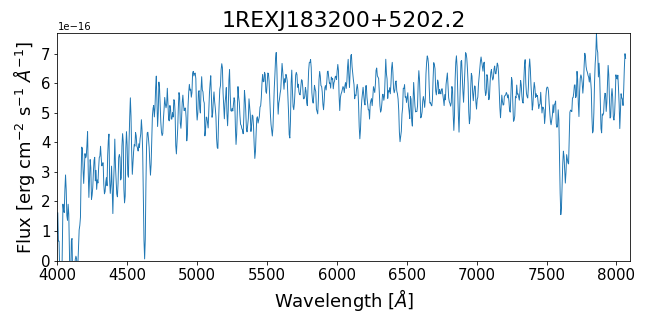} 
\includegraphics[width=4cm]{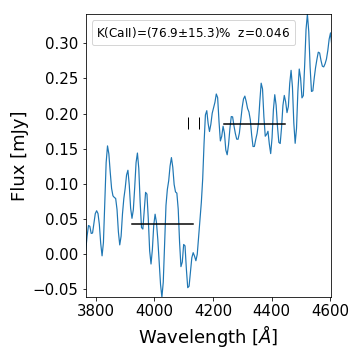} 
\includegraphics[width=9cm]{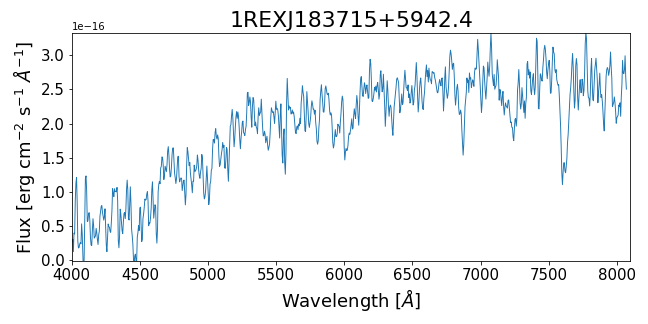} 
\includegraphics[width=4cm]{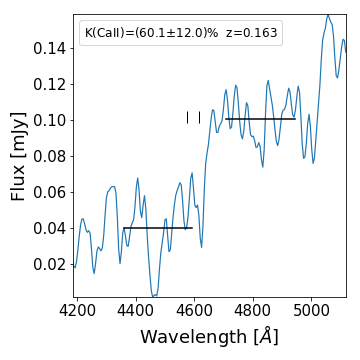} 
\includegraphics[width=9cm]{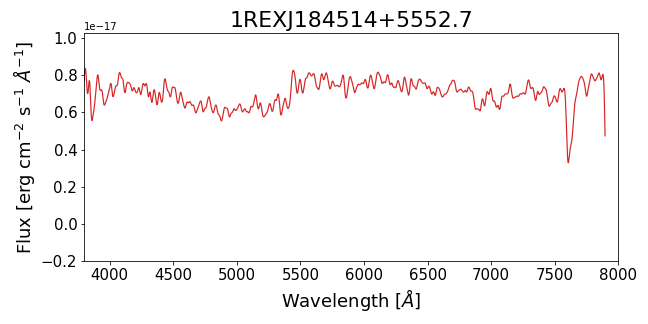} 
\includegraphics[width=4cm]{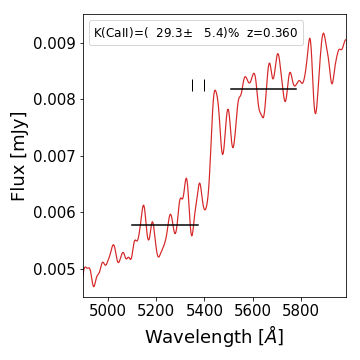} 
\includegraphics[width=9cm]{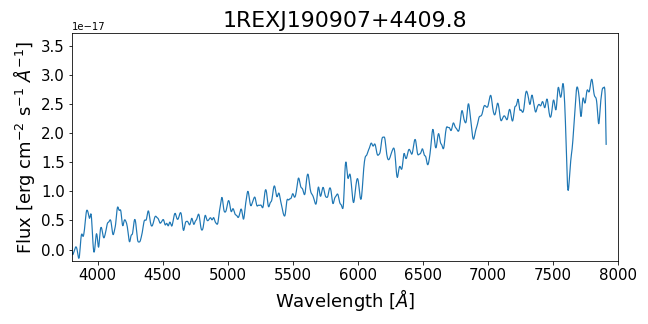} 
\includegraphics[width=4cm]{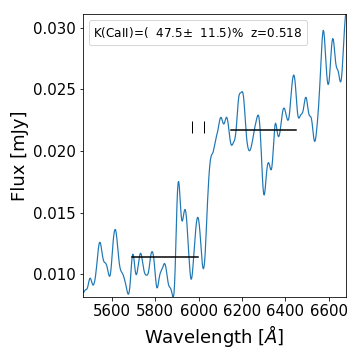} 
\label{nii}
\caption{continued. }}
\end{figure*}   


\bsp	
\label{lastpage}
\end{document}